École Polytechnique

MATHÉMATIQUES POUR LE TRAITEMENT DES MASSES DE DONNÉES

# *INTERNSHIP REPORT*

VERSION CORRIGÉE

*Submitted by*
Selma Mehyaoui
selma.mehyaoui@polytechnique.edu

*April-September, 2015*

INSTITUT DE BIOLOGIE DE L'ÉCOLE NORMALE SUPÉRIEURE
Laboratoire de neuro-éthologie du poisson-zèbre
German Sumbre

*How does neural activity encode spontaneous motor behavior in zebrafish larvae ?*

*Le corps du document étant rédigé en langue anglaise, l'italique primera pour le français ou le latin chiffres exceptés*

# Abstract

Motor behavior can be a response to a specific stimulus (in the sense that an agent perceives a sensory stimulation and moves as a reaction to it) or, sometimes, can occur without any obvious external causality (even in the relative absence of any stimulation, all animals including humans readily exhibit spontaneous behaviors).

Although most studies have considered spontaneous movements as a byproduct of fluctuations in neuronal activity, the generation of spontaneous behaviors is critical for exploration of surrounding environment, active sensing (the production of any signal by energy that would be proper to the fish, see for example echolocation for dolphins) or motor learning (how specific movement patterns are learned and allow to define a dictionary of specific movements).

The origins of spontaneous movements have been investigated in human as well as in other vertebrates. Studies have reported an increase in neuronal activity one second before the onset of a given movement: this is known as readiness potential. The mechanisms underlying this increase are still unclear.

Zebrafish *larva* is an ideal animal model to study the neuronal basis of spontaneous movements. Because of its small size and transparency, this vertebrate is an ideal candidate to apply optical recording methods. Moreover, zebrafish *larva* navigates the environment by producing discrete stereotypical tail movements, called 'bursts'. Nowadays, a large library of transgenic and mutant fishes is available, enabling us to target specific cell types or provide vertebrate models of various human neurodevelopmental and neurodegenerative diseases (Deo and MacRae, 2011) [12].

In order to understand what neuronal activity causes the execution of a specific tail movement at a given time, we will mainly use a prediction approach.
As A. Jouary would say, neurons whose activity anticipates soon enough the occurrence of a spontaneous movement "are good candidate for being neurons causing it".

# I. Introduction

According to the classical Stimulus/Response paradigm, we will consider a behavior as being internally driven when it cannot be linked to any external stimulus.
A typical and well known ambiguity between spontaneous and internally-driven behavior is thus possible, leading to a wide range of shades and possibilities between a purely hazardous and noisy electrico-neural field and a mysterious willing-fed non-decryptable neural process.
This distinction is not trivial : if I will give few clues in this introduction, it then won't be emphasized in the rest of this essay, voluntary limited to a mathematical analysis of zebrafish neural activations, before behavior and within the described experimental framework.

Under the Stimulus/Response paradigm, it is tempting to consider that the pathways involved during sensory-based decision making can be activated by neuronal noise causing internally driven behavior. However, such a case does not exhaust a hypothetical whole range of possible explanations. Internal drives such as hunger or anguish can give rise to "spontaneous" movements in the absence of immediate external stimuli, as if our own set of memories as well as our representational world were intelligible as parts of a great signs reservoir, full of auto-generated or auto-sustained "in potential" abstract stimuli, capable to cause movements and reactions without real visual nor auditory or olfactory stimulations. A kind of closed electro-physiological loop, auto-sustained by experience of its own memories and past reactions, agglomerating into something pedantly called "self".

I say it grossly but, for curious souls, no need for further comment since a whole field of Research do think these matters.
As I guess, cognitive sciences do propose wide and wild perspectives on the subject.

Although it is not trivial to determine the motivation underlying a spontaneous movement, several reasons have been advanced that show why internally driven behaviors should not be considered as the output of a noisy system, with no biological relevance. What's more, driving force of internally driven behaviors can be casted into two categories: extrinsic or intrinsic motivation (Gottlieb et al., 2013) [9].

In extrinsically motivated contexts, behavior is a [illegible] to reach a biological goal e.g. finding food or potential mates. The goal of internally driven behavior is not to act on the environment but to retrieve information.

Thus, the distinction between extrinsically and intrinsically motivated spontaneous behaviors is not always easy to draw : it however illustrates with a sufficient level of evidence, and of "epistemologically valid' "distinguishability" (such binary classification both makes sense and can be useful), that distinct motivations can underlie internally driven behaviors.

In the present work, we thus intend to find mathematical methods that permit to predict self generated behavior observing the dynamics of the brain, with this *a priori* general knowledge we have about the idea of "self" (first dictionnary question) "generated" (second dictionnary question) "behavior" (third one for this sentence). Here again, for any semiology-oriented question, both hard sciences and Philosophy should help in the limit of their common power of nomenclature, which is as I know a shared tresure for all of us.

To be perfectly rigorous, before any further proposal, we first should be able to show that activity preceding a motor behavior is systematically and significantly different from activity observed during inertia.

Without such a basic first insight, the very heart of this internship is in itself made.

Since we have a large neural population (⇡ $10^4$ neurons), we will have to use tools belonging to High Dimensional Statistics to solve such a challenge. We don't focus on pre-motor area since we know that neuronal activities related to decision making are widely distributed (Cisek, 2012) [11].

What's more, the number of spontaneous movements is relatively low compared to the duration of the recording, so we face a highly unbalanced problem : we will have to adapt our methods to such a specificity. At the end, we will propose some labeling for tail movements in order to see if they can be predicted more specifically given neural data.

Concerning the philosophical aspects sooner raised or evoked, the reader should easily find a dense and constantly enhanced literature on the subject, if libraries do still exist when and where these lines are read. Less obviously, he could also catch the attention of a honest and devoted Philosophy teacher if no replaced by a bot (which could so be the maybe-great causal creation of a surely great causal mind), and that in order to sharpen his thinking on this tough problematic.

What I clumsily mean by these few lines is that one can be fully aware of and deeply moved by this whole dedaleous questioning, without willing to convoke both philosophia and an heuristical scientific construction in the same paper : so, since it's not the "topic" here and since a huge and venerable amount of work is proposed on the new and wise scholastic field of 'neurophilosophy', let's dare go on our modest journey.

I will present in a first section the acquisition and pre-processing methods for both neural data and behavior data (tail movements movie).

Then, I will give an insight of the methods built and coded during this internship to analyze this material.

I will conclude comparing the outputs of these different explorations and giving an idea of what could be done next.

# II. Method

We are using selective plane imaging microscopy to record from large population of neurons across multiple brain region while recording the spontaneous motor activity of the *larva*, ⇡ 10000 neurons are recorded 2 hours during at 10 Hz.

## II.1 Experimental setup: SPIM analysis

The technique permitting to register both neural activity and tail movement of the fish is based on two tools: one is genetic, and allows to make *larvae* neurons be fluorescent when firing, the other is optical, and consists in the whole setup that permits to observe in real time both this neural activity and fish motor behavior.

### II.1.1 GCaMP5G zebrafish

GCaMP5G is a genetically encoded calcium indicator protein: by extension, we call GCaMP zebrafish a fish that has been genetically modified such that it expresses this fluorescent protein. As a consequence, as calcium takes part in the ionic transfer of any neuronal transmission, when a neuron is active it becomes fluorescent; so such a fish's brain activity can be recorded through neural imaging, lighting his brain with a proper wavelength lightsheet.

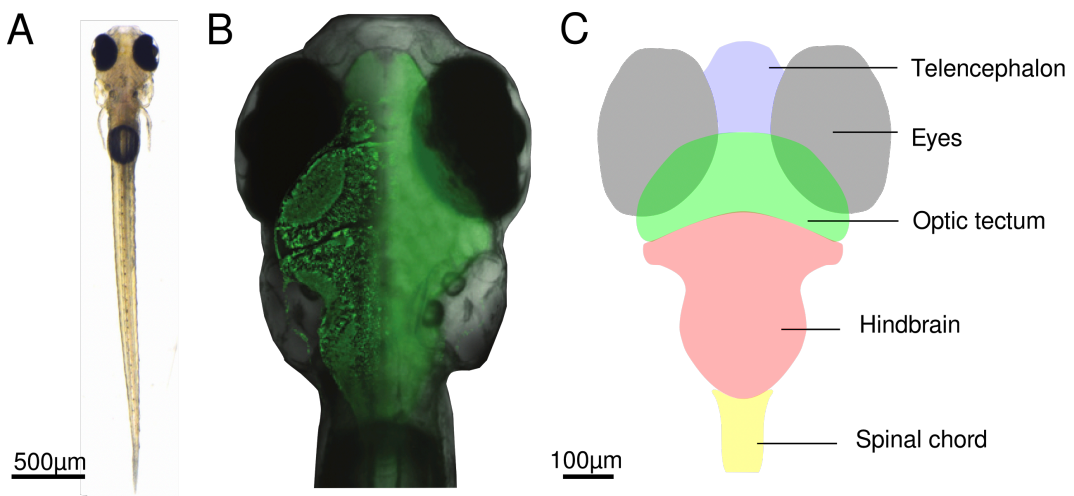

Figure 1: Coarse brain anatomy of a 6 dpf zebrafish *larva*. (A) Bright-field image of a zebrafish *larva*. (B) Overlay of a bright-field image of the *larva* head with images of its brain acquired using two-photon microscopy (left part of the brain) and fluorescence imaging (right part of the brain). Note the spatial resolution on the left part obtained with a two-photon microscopy. Neurons are labeled with the green fluorescent calcium indicator GCaMP5G. Image reproduced from Fetcho (2012). (C) Schematic drawing of the *larva*'s brain showed in B representing the main parts of the brain (telencephalon, optic *tectum*, hindbrain and spinal cord) and the eyes. The 100 micrometers scale bar is common for B and C.

### II.1.2 Optical setup

#### Neural activity recording

The setup used in the lab to register data is selective plane illumination microscopy (SPIM). A lightsheet lights a plane of the brain whose activity is recorded by a camera. The head of the fish is stuck in a fixed position of the plane thanks to agarose gel. SPIM set-up is a part of the whole acquisition tool presented in Fig.2. The record is made at 10Hz.

#### Behavior recording

Tail movements are recorded to, thanks to another camera device, this time at 200HZ : the device is also presented in Fig.2, as another part of the total set-up.

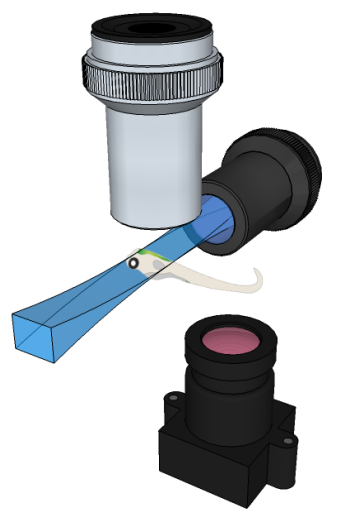

Figure 2: SPIM setup corresponds to the top camera and the lightsheet device, behavior is recorded thanks to the camera represented below the tail.

## II.2 Data pre-processing

Once we have collected both brain recording and tail movement videos, we must pre-process the data in order to construct exploitable variables. The goal is to be able to construct the following two objects :
-X will be the matrix corresponding to the neural activity.
-Y will be the vector indicating if there is a movement or not for the selected observed times.

Once we have the movie, we don't process it pixel by pixel but ROI by ROI (when ROI means Region Of Interest). Ideally, ROI should be a neuron. Here we work with a formerly developed tool for neurons detection, which grossly detects ROIs that correspond to neurons or groups of neurons. But in later work we will work with mega-pixels, since there is no loss of information if these hexagons are small enough compared with the mean size of one neuron.

Then, at each time, for N ROIs by image, we have the given of the fluorescence associated with these groups of pixels. The relevant data is the ratio of the difference between fluorescence value and the average fluorescence upon the average fluorescence $\frac{\Delta F}{F}$.

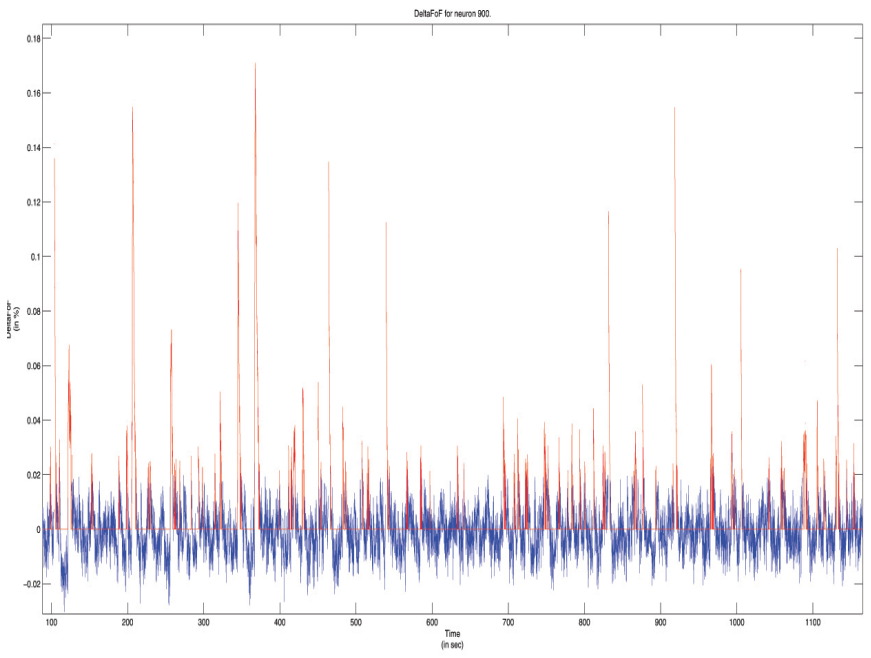

Figure 3: Thresholded (in red) and rough (in blue) activity for neuron 900. That signal can typically be seen as one line of a raw raster.

What's more, once we extracted this data we apply a threshold in order to keep information related to picked activity. That allows us to be sure we don't work with neural activity relative to noise. Activity that is under the threshold is put to zero when recorded into an analogic or binary raster.

Here are the two raw rasters (before discriminating between times that correspond to movement occurrences, times before movement, and inert times).

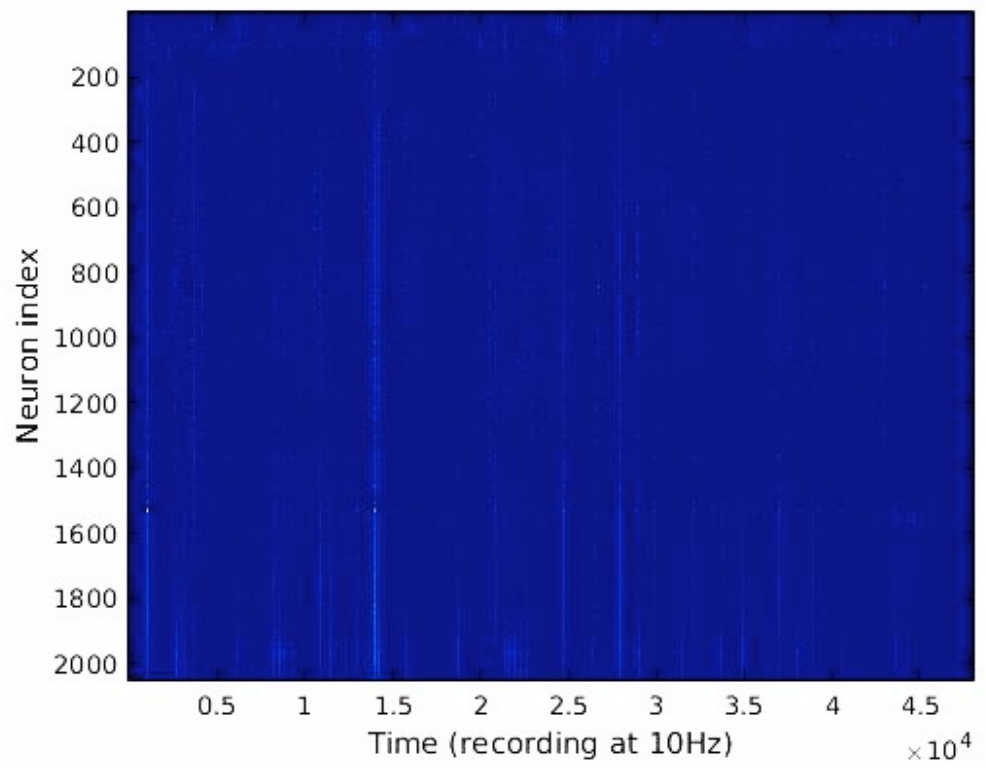

Figure 4: Row analogic raster plot. Each line represent one neuron's activity during the whole recording time.

The first one is an analogical one, the second one corresponds to the binary raster plot.

A simple thresholding on the first matrix is sufficient to obtain the binarized one.

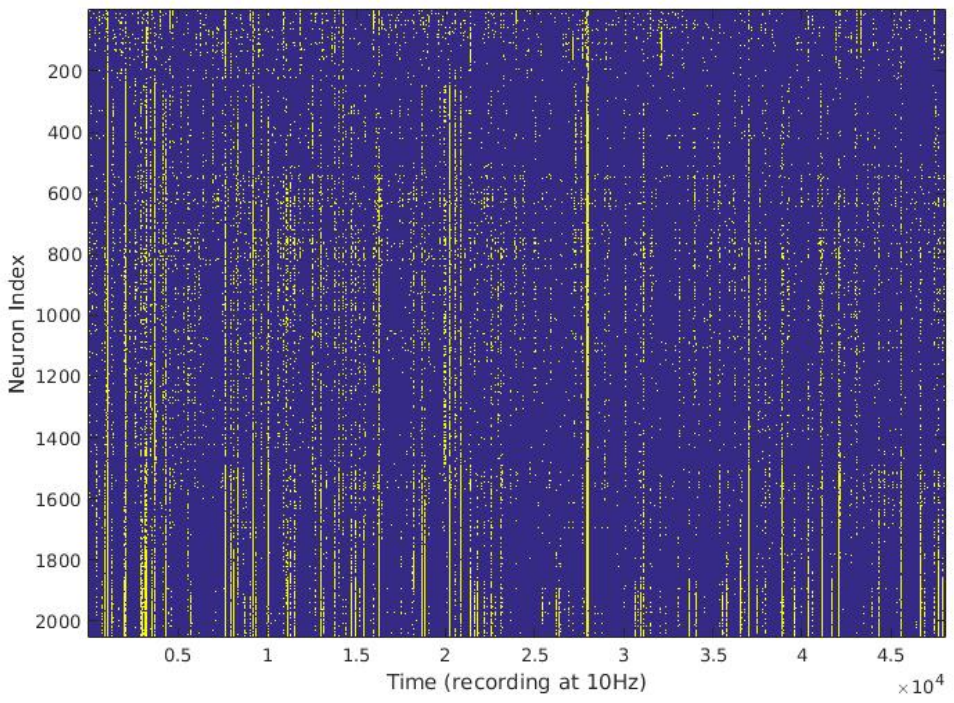

Figure 5: Binary raster plot: applying a threshold to the analogic raster plot showed above, we obtain a binary raster.

From now on, we will consider that each ROI is a neuron.
The definition of a movement's occurrence needs a thresholding of tail movement data in order to distinguish "noisy" oscillations of the tail from real movements.

This work takes part in the data pre-processing framework as an important step for defining what a "movement" is within our heuristic try, and what an inert time is.
For that, we threshold according to the energy of a movement, and according to its amplitude (details should be asked to A.Gramfort, to whom I had sent a report on the subject).

Here is a caption of movement selection from rough tail movement recording:

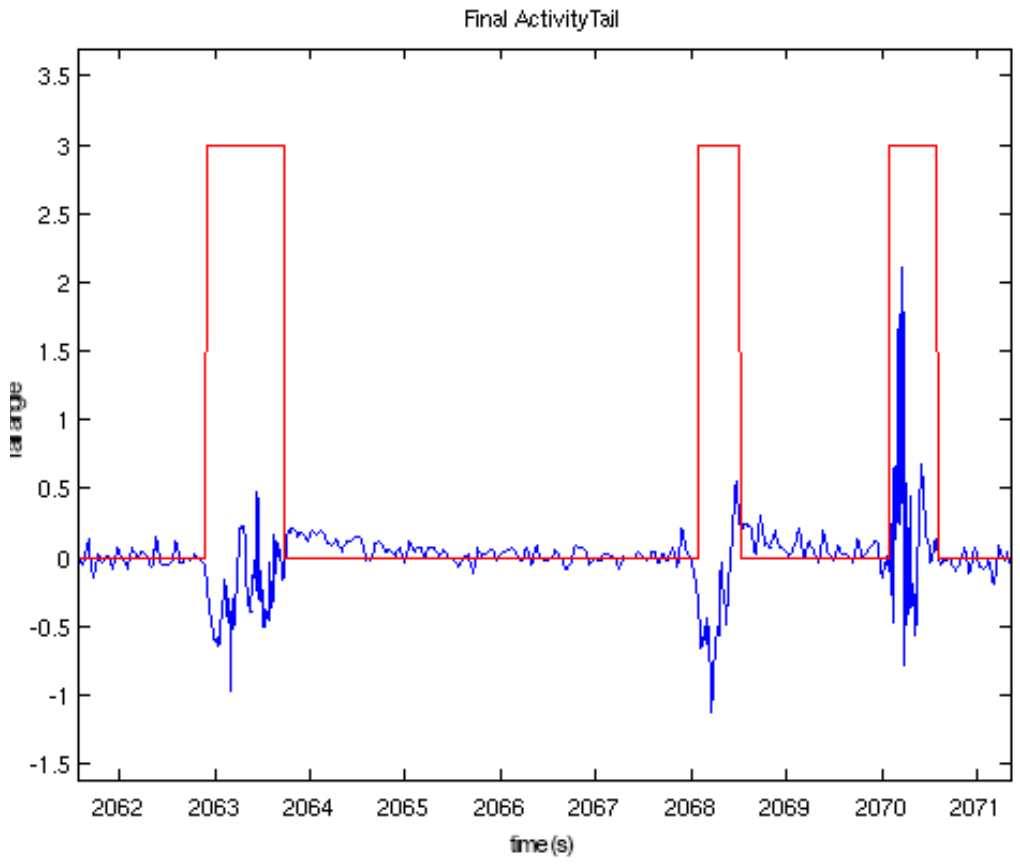

Figure 6: Tail movement signal filtering will allow distinction between movement (and preparation of movement) times and inert times, and then Y vector construction.

Despite agarose gel, the head of the fish moves during movement: that's why a raster cannot be defined while a movement does occur. Therefore, we can observe neural activity during preparation time of movements and during inert time (times during while no movement happens).

What's more, we want to know how/if a movement can be predicted thanks to neural activity before it. In this purpose, we should be able to compare neural activity that precedes a movement and neural activity that is far enough in time from the occurrence of any movement to be considered as not correlated with behavior generation.

As a consequence, after having detected all movements occurrences, we keep the frames of all neural activity recorded between 0.1 and 3 sec before any movement as 'Preparation' frames ('Prep frames'), and all the frames that contain activity that is more than 6 sec far away from any ending of movement or any beginning of Preparation frame (*i.e.* movement's beginning 3sec) as 'Inert' frames.

So, on one hand, for each one of the movements, we have the associated preparation activity.

On the other hand, a set of 'Inert frames' that give several 3sec window recordings of neural activity, considered as describing what happens in the brain while no movement is done nor prepared.

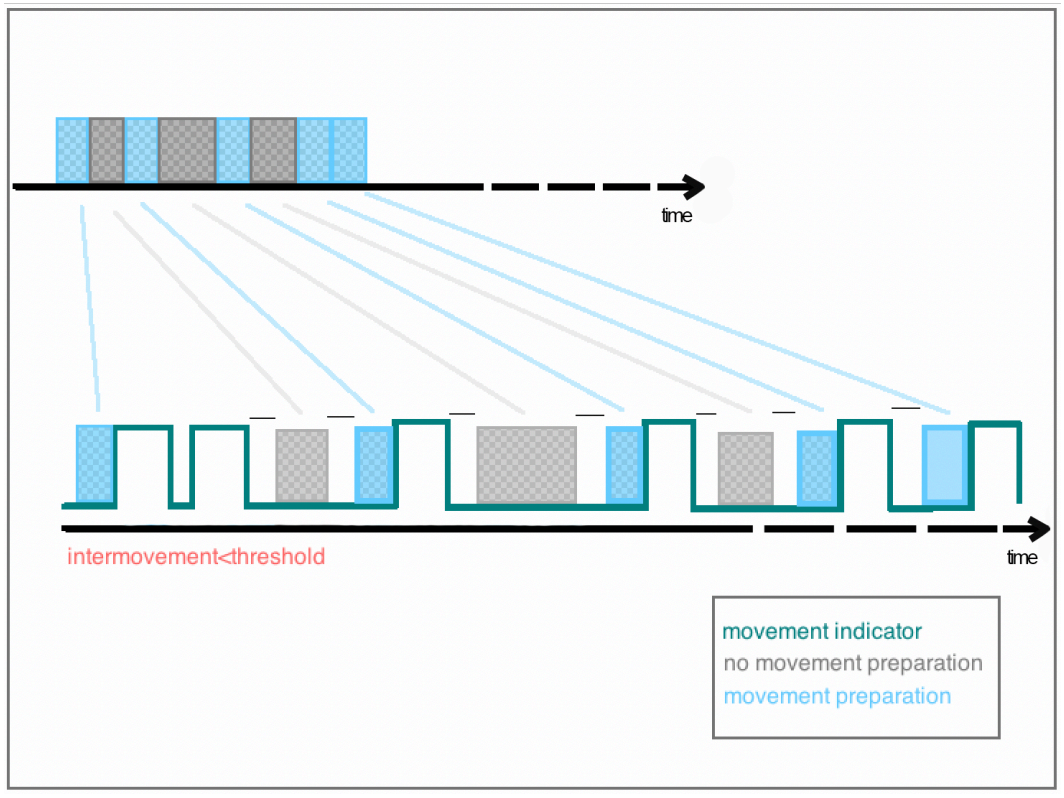

Figure 7: Signal filtering and frame construction.
On the top the frames extracted from row signal, on the bottom how we decide that/if a time point belongs to an inert frame (in grey) or to a preparation frame (in blue).
During movement (movement indicator is plotted in green) the head (and so, the brain) moves: thus, neural recording cannot be kept, whereas before movement (blue frames) it can be recorded as 'movement preparation's activity' (Prep frames). When there is no movement, and keeping in mind that we don't want to be too close to the beginning or to the end of a movement, we keep recorded activity as 'inert activity' (Inert frames).
We impose a delay between movement occurrence window and inert times that we keep, in order not to record the decay of neural activity that corresponds to movement generation in inert frames.

The given of $\frac{\Delta F}{F}$ for every neurons, at every time $t \in [1, T]$, is recorded in what we call an analogic raster plot. It will be our entry variable, X, as it gives neural activity across time.
If we average the signal for each frame (prep frame or inert frame) of $[1 \ldots N_{frames}]$, we then work with what we call an average raster.

To sum up, we can work with:
-X as an analogical raster recording activity during $N_{frames} \times T$sec where T is the length of recording we keep for each frame 1.5 or 3 sec depending on the analysis). That means that we could see each line (one line per ROI) as a concatenation of a thresholded analogical signal $\frac{\Delta F}{F}$ as presented in fig 1.
-$X_{avg}$ as an averaged raster where each line is the concatenation of $N_{frames}$ values of a mean value of the thresholded $\frac{\Delta F}{F}_{t_{frame} \in [1 \ldots T]}$.
-$X_{bin}$ as a binarized version of the analogical raster: each value above the threshold is put to one, all other to zero.

The output variable, Y, is a binary vector of length $N_{frames} \times T$ or $N_{frames}$ (depending on the fact that X is averaged window per window or not) indicating if the frame is a preparation frame or a frame corresponding to an inert time.

Figures are small and, in the raster plot case, sometimes hardly readable in this document. The code is nonetheless simple to re-implement once similar *data* have been obtained.

An interesting observation consists in underlying that inert times do not imply lower activity, but maybe less discriminant activity (which is not the same, see for example the vertical whole-population activations in the inert-frames raster). This corroborates the idea of a fine and qualitative coding which would encode the information of an incoming movement by a sparse and subtle combination (or coïncidence) of activations.
Also, this same answer will be done to anyone (legitimately at first glance) surprised by the presence of whole-firing patterns inside recorded activity for inert-times raster.

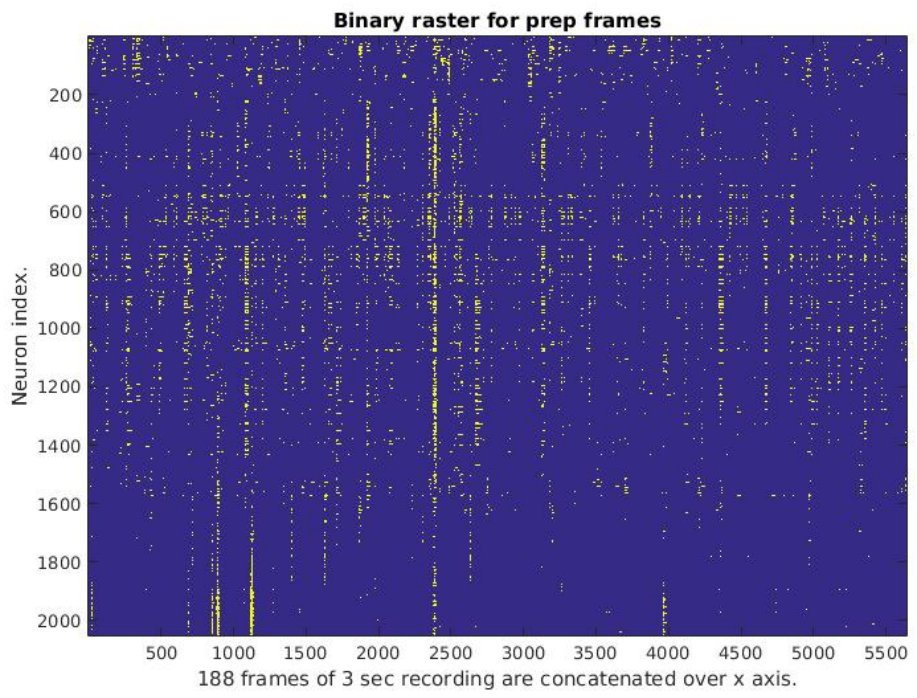

Figure 8: Binary raster for prep frames. x axis refer to time. All 3 sec recorded frames selected as coding for movement preparation have been concatenated. Each line of the y-axis corresponds to a ROI/neuron. Distinction between prep frames and inert frames is made thanks to the processing presented in Fig.7.

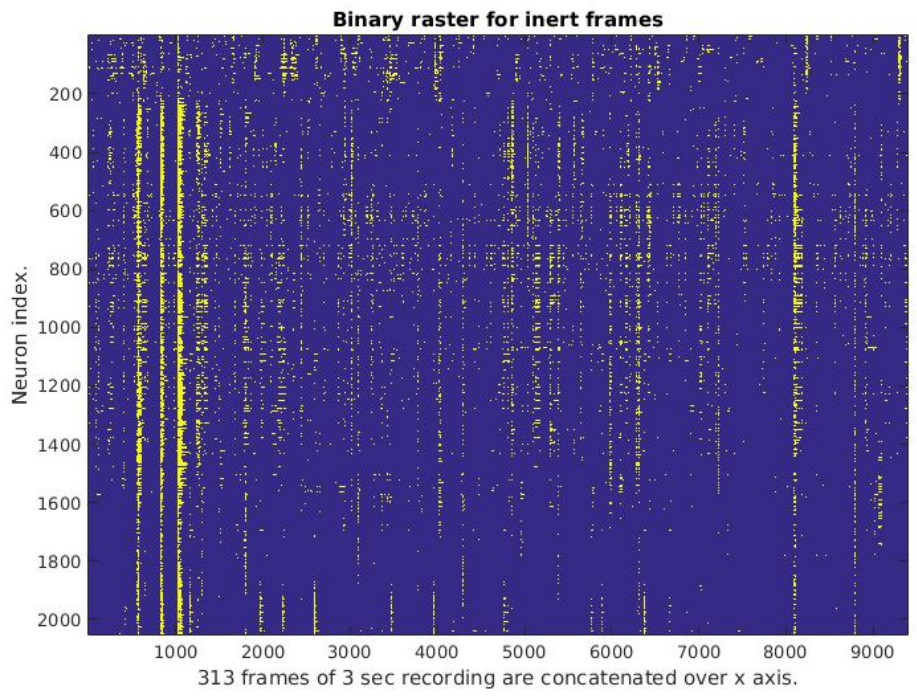

Figure 9: Binary raster for inert frames. x axis refer to time. All 3 sec recorded frames selected as referring to inert times (no motor activity) have been concatenated. Each line of the y-axis corresponds to a ROI/neuron. The distinction between prep frames and inert frames is made thanks to the processing presented in Fig.7.

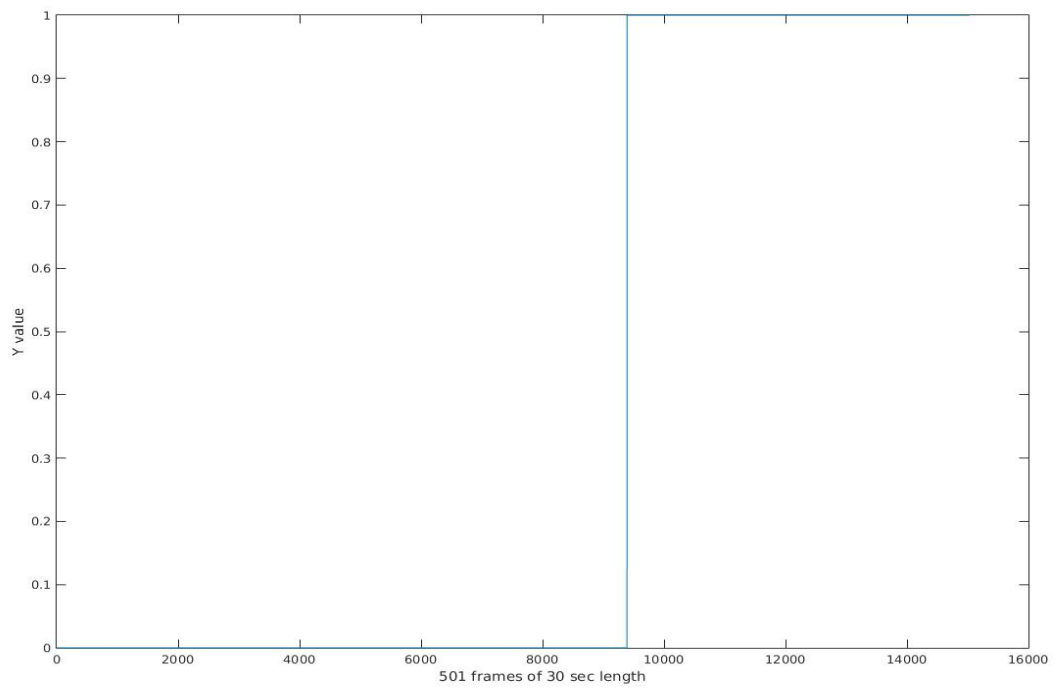

Figure 10: Y is the movement indicator, at zero for the $N_{inert}$ first frames and at one for the $N_{prep}$ last frames. If we wanted to classify movements into different kinds of motor behavior, Y would be a label vector.

# III. Results

## III.1 Descriptive approaches

We first focus on what we call descriptive approaches: that means that they don't extract any information about how the movement is caused, in fact there is no causality information extracted by these methods, but just a first intuition of how predictive some neurons could be, with no more details. In other words, we do not implement here any regression model, we simply use really basic mathematical stuff to see how neural activity changes in function of the occurrence or not of a movement.

### III.1.1 Univariate statistical test

We compare, for each neuron, if its average activity on frames unrelated to motor preparation or execution (inert), is different from its average activity on frames preceding a movement.

We first tried a t-test and then chose to use a Mann-Whitney test.

Let $Y \in R^{N_{frames}}$ be the binary label vector.

$Y = [Y_t]_{t=1..TxN_{frames}}$ such that:
$\forall t \in [1..T \times N_{frames}]$
- $Y_t = 1$ 300ms or less before movement,
- $Y_t = 0$ during inert frames.

As a consequence, we can also define categories for neural activity (pre-movement neural activity, inert times neural activity).
- The set of the inert frames can be defined as: $X_0 = \{X_{avg}(t,:) | Y_t = 0\}$.
- The set of the pre-movement frames can be defined as:
$X_1 = \{X_{avg}(t,:) | Y_t = 1\}$.

Then, we made the test run for two hypotheses:
- H1 : $E(X1) \geq E(X0)$. The activity of neuron $j$ recorded in $X_{avg}(:,j)$, is greater before a movement than during inert times.
- H2 : $E(X1) \leq E(X0)$ . The activity of neuron $j$ recorded in $X_{avg}(:,j)$, is lower before a movement than during inert times.

We get two p-values maps, one per hypothesis.

- p-values associated with hypothesis H1 will be called *PvaluesEx* (in reference to the excited state, but it is clearly a non-valid language shortcut).
- p-values associated with hypothesis H2 will be called *PvalueIn* (here again, it is a non-valid shortcut referring to the word 'inhibited', mechanically invited here and definitively avowed as a fault).

<u>T-test</u>

On H1 map, we recognize zones known as involved in induced activity movements: this mapping can be seen as a first indication of implied zones for the preparation of spontaneous movements. The map of H1 shows higher activity in the hindbrain, the last anatomical relay before the spinal cord.

What's more, we observe that p-values evolve in two radically different orders of magnitude depending on the test ($H_1$ or $H_2$).

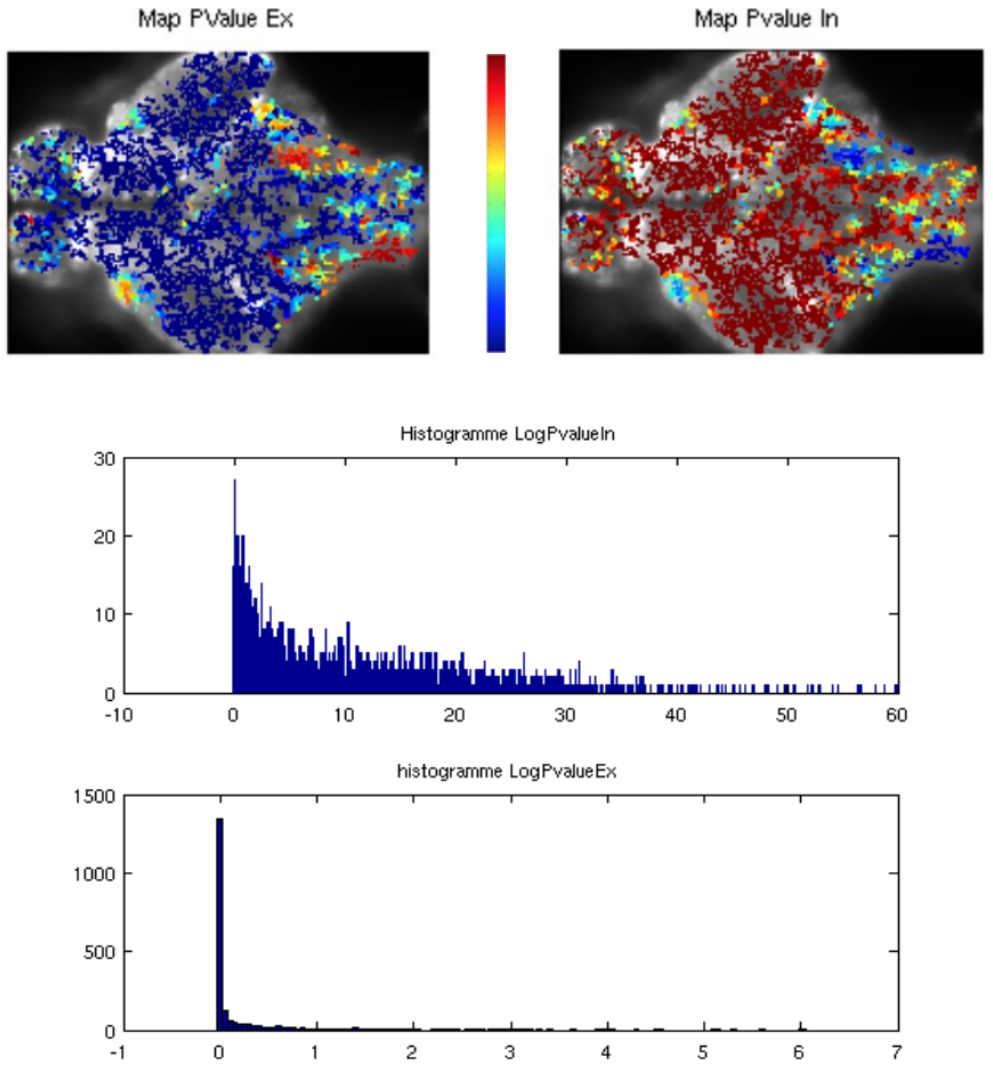

Figure 11: Map and histogram of associated p-values for t-test on H1 and H2.

<u>Mann-Whitney test</u>

Mann-Whitney test is more adapted as there is no assumption made on the distribution of the data. The range of p-values is more reasonable using the Mann Whitney (up to $10^{5}$) than the t-test ($10^{14}$).

The relative fluorescence is thresholded, thus its distribution is highly peaked in 0, the gaussian hypothesis of the t-test is no valid.

Concerning the map, we don't see any significant difference with t-test map. If we look to score values, the range is less spread for $H_1$ less packed around zero for $H_2$.

A surprising result is the zone close to the optic tectum, and usually associated with stimuli response, that exhibit pretty good score for $H_1$ in both t-test and Mann-Whitney test.

This preliminary map allows us to expect that dynamics of prediction could be found in other zones than pre-motor area (that are massively represented in both maps, and, since they refer to neuropil, do not allow neuron-resolution interpretation).

Below, the results we got.

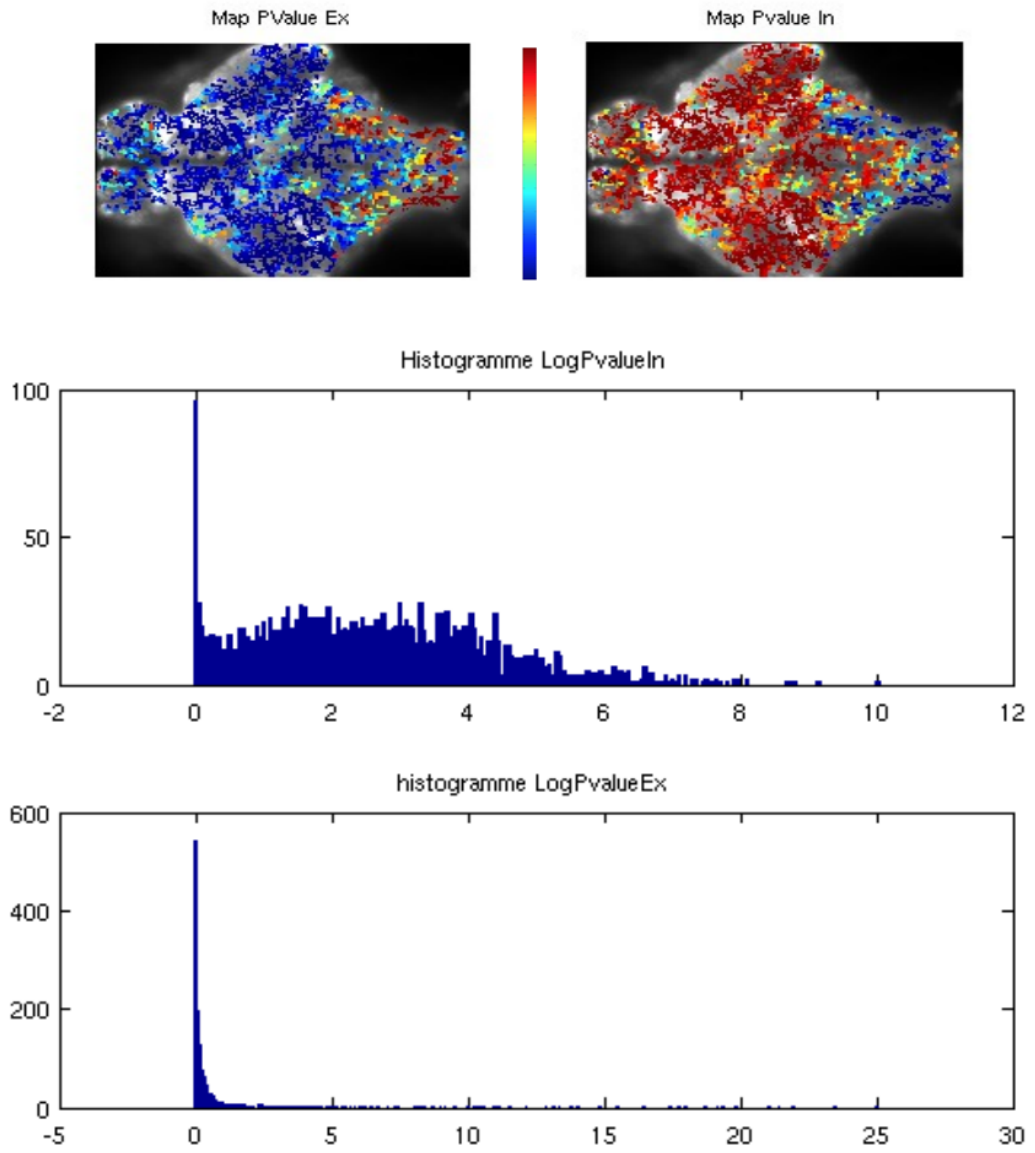

Figure 12: Map and histogram of associated p-values for Mann-Whitney test on H1 and H2.

As a conclusion, we can say from these tests that our *data* set seems exploitable for further analysis, since we do not get any aberration for such a simple insight on their movement prediction ability.

### III.1.2 Searchlight

Searchlight is a commonly used method used in fMRI to extract topological features from brain activity *data*. The brain is scanned by a spot (a small spherical subset of neurons, called "searchlight") that extracts from the activity of targeted neurons the score of a support vector machine for predicting the occurrence of a given event. This score tells us how informative the spot is.
We first adapted this method in zebrafish *larvae* using spots of 30 neurons and using a polynomial kernel support vector machine to predict the occurrence of a movement given the average neural activity on a 1.5 sec duration time before movement.
Such an approach will give an insight of how informative some regions can be, or how close to such a simple classifier model the prediction can be modeled.

#### How it works

At each neuron location, we construct the spot $S_j$ such that $S_j$ contains $j$ and its 30 closest neighbors. We then train an svm spot by spot, using inner cross-validation (60 folds) to estimate its performance on balanced *data* (class 0 and class 1 are balanced and add up to 400 examples in the training set). At first, this performance is recorded in $h_{test}$, which is the average of correct label obtained on each fold k of the cross-validation.

We chose a quadratic kernel after having tested several different polynomial kernels: its performances are better than a linear kernel (which give a maximal score of 60% against 66% with the second order), and incrementing again the order of our polynomial does not significantly improve its performance. We have visualized the *data* in a 2 dimensions space thanks to a PCA, in order to have better intuition of the relevancy of such a kernel. This choice seems confirmed as good enough to give us correct prediction scores. The location of the most predictive searchlights does not change with the choice of kernel.

Before mapping the results, we use Platt-scaling to enhance interpretability, and manage the imbalance of the classes. Platt-scaling plot a probability distribution on the class, and allow to eventually adapt the offset of the plan in order not to suffer from the imbalance of *data*.

Calling f(x) the distance of a point x from the discriminative hyperplan defined through the svm, Platt scaling expresses:

$$P(y = 1|x) = \frac{1}{1 + \exp(A.f(x) + B)}$$

where A and B are computed through a maximum likelihood computation, on the same training set that has trained the classifier.

Thanks to such a transformation, we can translate the separating hyperplane in order to compensate an eventual overfitting and then look at the score as the area under the ROC (the ROC curve is constructed by continuously changing the offset of the hyperplane. The number of folds necessary for the cross validation had to be tested looking at the variance of the scores for each fold according to the chosen number of fold. We came to the conclusion that at least 400 folds (with, each time, a 6 frames test sample) were necessary to have reliable scores. This exploration phase took a lot of time to run, since we had to make the algorithm run on our *data* a lot of time for all the possible cross validation partitions. After having fixed all these parameters, we got results with performances that go from 0, 4 to 0, 66.

#### Maps

Finally, we map these performances (we chose the area under the ROC as an efficiency indicator) spot by spot. As there is one spot per neuron (each spot is centered on one particular neuron) this gives a map that seems to have a neuronal resolution but that must be read as a grossly drawn idea of prediction power of brain at a 30 neurons region resolution:

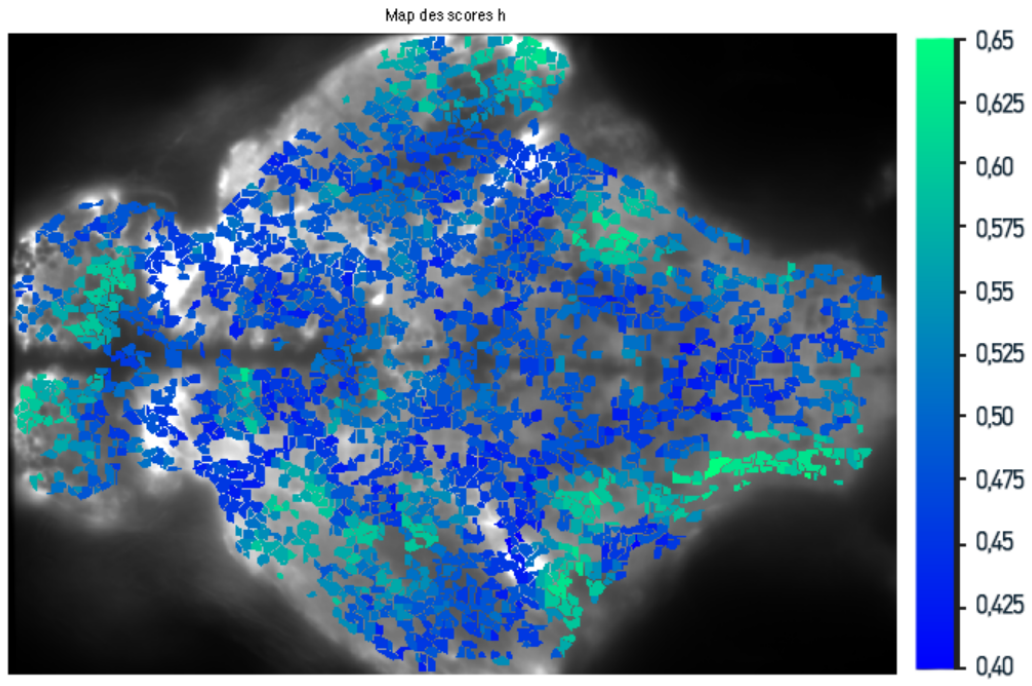

Figure 13: Components 1 and 2 for method B version of PCA.

The distribution of the scores is given below.

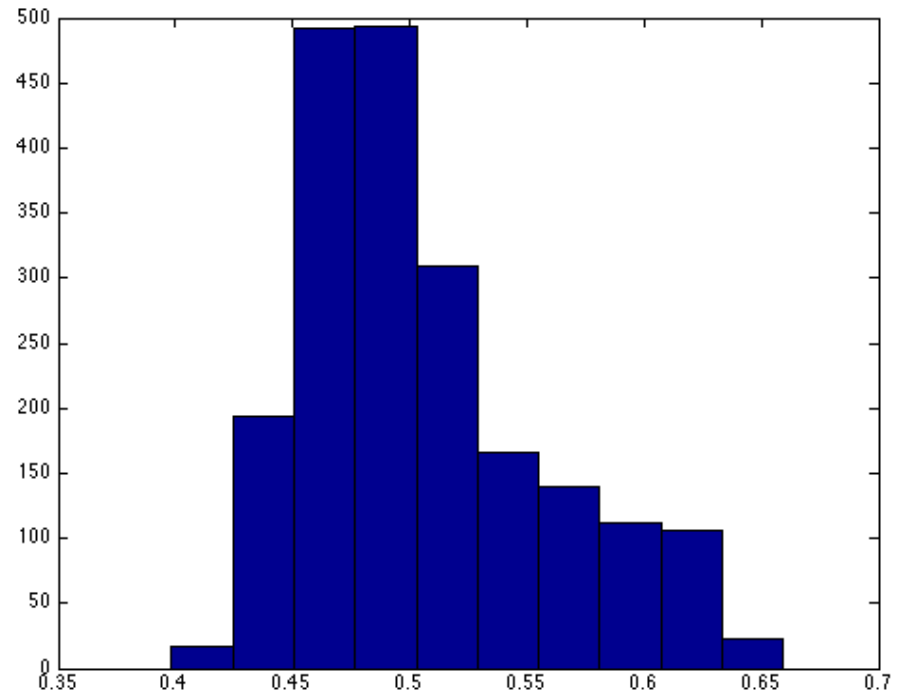

Figure 14: Distribution of the scores

<u>What does it tell ?</u>

We get coherent results with the first statistical description (Mann Whitney test), in the sense that we find zones that had been exhibited by these tests. But we see here also neuropil activity, captured by svm classifier as an activity good for prediction. The problem is that we would like, at this point, to have a more precise given of neuronal zones involved, instead than an high view of active zones that we know already as involved in movement generation (it is the case for neuropil for example).

That's why we cannot keep these results as sufficiently relevant ones for further analysis, even if the method suggested by the searchlight can have interesting application if we adapt it to a dynamic approach, or to a labeled-*data* prediction.

Such a customization of the method will be proposed in later work.

## III.2 Dynamics of prediction

After these two first grossly descriptive approaches, we want to focus on the temporal dynamics of movement's predictability. To do so we have two solutions :

• We can analyze the whole analogic $\frac{\Delta F}{F}$ captured for all t in [100:100:1500] ms before movement, that means that we would work on a vector of length 15 instead of working on the average values as for the statistical tests. Principal Components Analysis or Independent Components Analysis can then be use to find out temporal or spatial patterns.

• Instead of working on average values of neural activity before each movement, we work time by time. On other words, we run an analysis for all frames (prep frames, inert frames) and, for each one of these frames, we take $\frac{\Delta F}{F}$ value at a specific time t in [100:100:1500] ms before movement.

<u>III.2.1 PCA</u>

Principal Component Analysis is a well-known method used to find a basis of patterns in which the signal can be reconstructed with a controlled error. The advantage is that the covariance between each pattern is minimized.

Here, I chose not to perform a PCA directly in the analogic $Raster_{prep}$. Indeed, we are looking for patterns that are relevant for any frame before movement, but we do not care about what happens during inert frames for constructing the basis.
In this part we only consider a sub-sample of X such that only class 1 (Y = 1) is represented (that is we focus on pre-movement frames).

Let X1 be the pre-movement neural activity matrix. $X_1=\{X_{i,t_j}|Y_j=1\}$.

I tried two approaches (Fig.15):

• The first one (Method A) consists in:
– first, applying a Principal Component Analysis to X1,
– then, constructing the average pattern on a window (*win*) of 2 seconds for each one of the components (*win* is the minimum duration of preparation movement observation).

• The second one (Method B) flips these two steps:
- first, we compute the average time course before a movement, for each neuron,
- then, we run a PCA on the average time course across neurons.

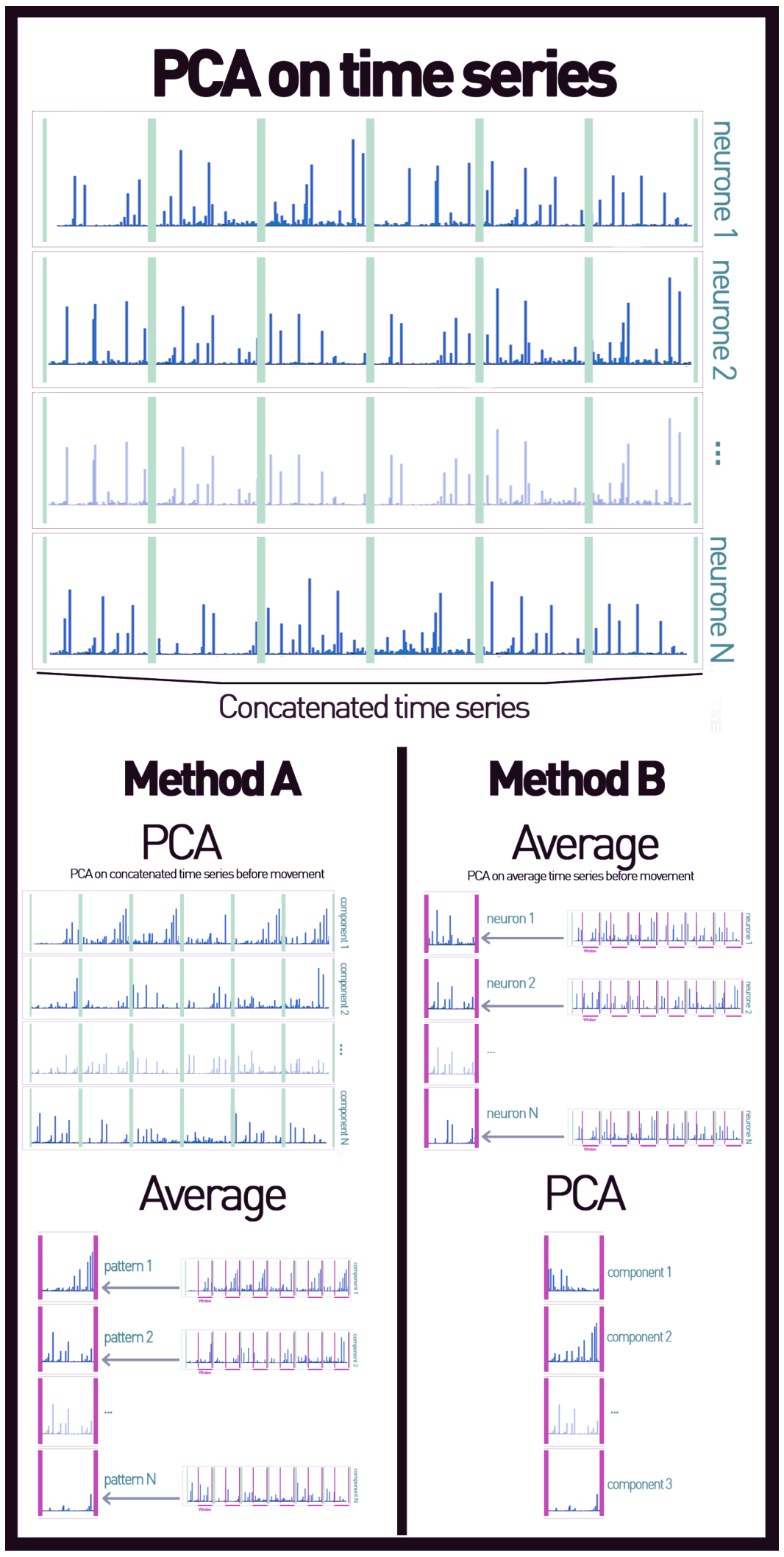

Figure 15: Two frameworks were imagined for adapting Principal Component Analysis to our *data* and our problem.

Looking for temporal patterns

I tried both method A and method B, and finally decided to keep method A as a reference method for this part. Let's first have a look to the components found by method B and let's explain why we cannot exploit these results.

I propose to have a look to the two next figures, that illustrate both temporal and neural maps that method B allowed us to extract.

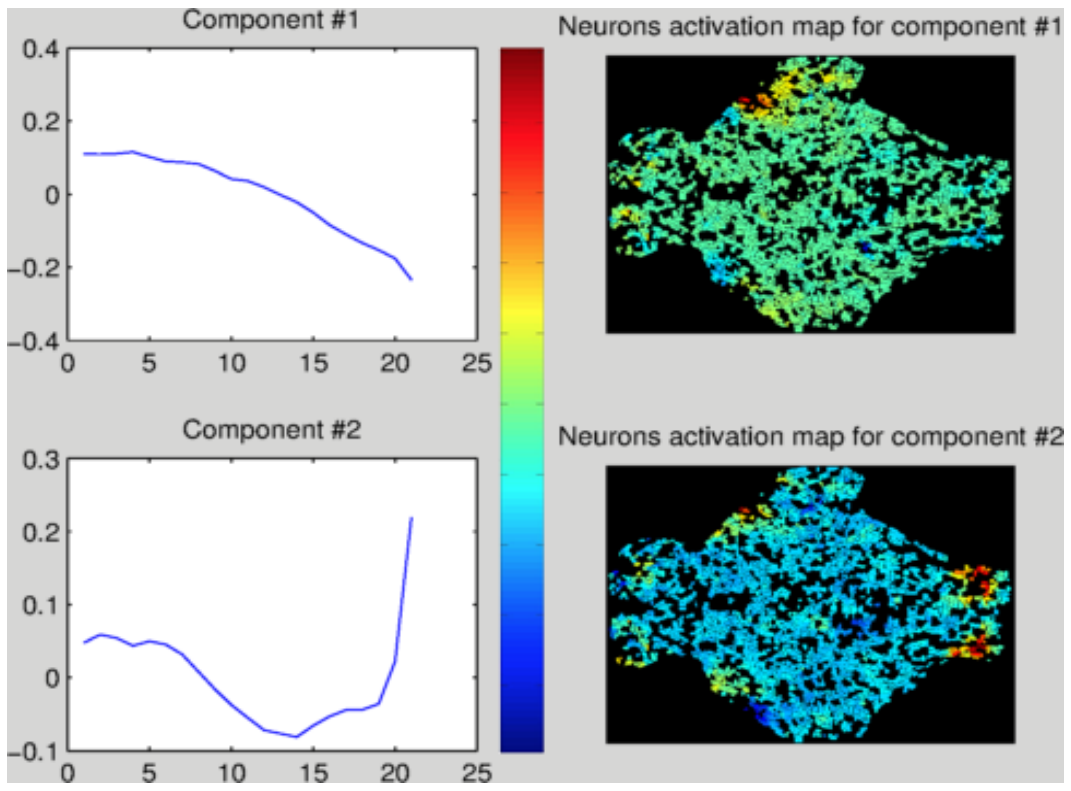

Figure 16: Components 1 and 2 for method B version of PCA. On the left, temporal dynamics on a 3 seconds window, on the right, the associated map of the component. We get for the two first components time tendencies that we will also obtain with method A, since there is a ramp like signal for regions that are corresponding to pre-motor area and a decrease for a zone close to visual regions.

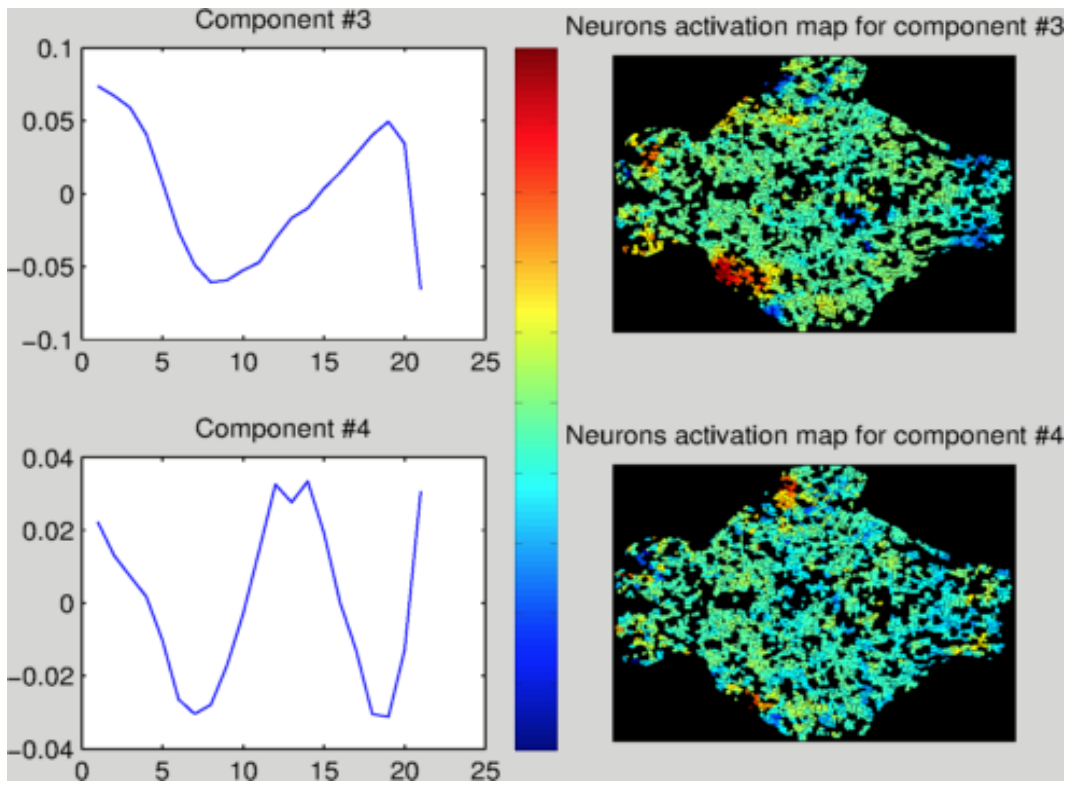

Figure 17: Components 3 and 4 for method B version of PCA. On the left, temporal dynamics on a 3 seconds window, on the right, the associated map of the component. These components allow to shed light on a problem we could have anticipated: indeed, we see for these components sinusoid like oscillations that indicate that we might face a problem of artefacts generation.

As we look to the patterns found by method B, we see that such an approach creates artifacts by averaging before the analysis: if there are small offset between two hidden dynamics, then by averaging before doing the PCA, this offset will be captured as a sinusoidal component in the PCA result.
This is a Fourier-like artifact.
Below, an illustration of this phenomenon.

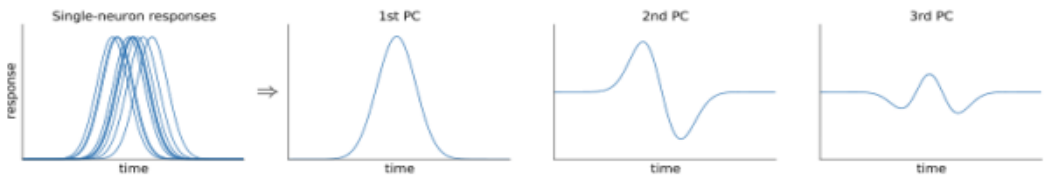

Components found by PCA (meth A)

From now on, we consider that we work in the framework defined by method A, *i.e.* that we first process an PCA and then average the signals obtained to get a specific pattern associated with each component (represented on a map).

To speak more formally, let's link each one of the variables obtained to formal PCA decomposition.
Given the analogic raster corresponding to the concatenation of activities corresponding to prep frames $X_1$, we perform a PCA to $X_1 - \tilde{X}_1$ by writing:

$$X_1 - \tilde{X}_1 = \sum w_i . \sigma_i . v_i^T$$

Where, $\forall\, i \in [1 \cdots N_{neurons}]$:
- $w_i \in K^{T \times N_{frames}}$ are the score vectors
- $\sigma_i \in K$are the latent matrix coefficient (*i.e.* the eigenvalues of the decomposition)
- $v_i \in K^{N_{neurons}}$ are principal axes
- $c_i = X_1.v_i = \sigma_i.w_i$ are the principal components, which express the coordinates of the lines of $X_1$ in the basis $\{v_i\}_{i \in [1 \cdots N_{neurons}]}$

In our cases, each line of $X_1$ represent the distribution of neural activity across neuron at one particular time.
That mean that our components will be a time signal, and our associated basis of representation will be represented as a map (we will plot the associated scores for each component).

Here are the results we got:

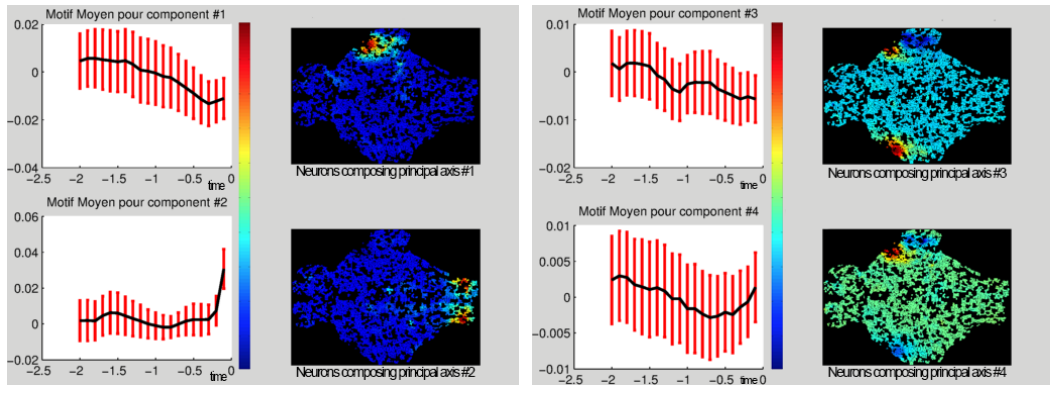

Later explorations will focus on the two first components, since they refer to zones that had been lighted by first searchlight results, correspond to well shaped ascending or descending ramps, and do correspond to the most weighted basis vectors for whole neural signal reconstruction (results are sorted by decreasing order of associated eigenvalue).

We see here that neurons close to optic tectum tend to have a decreasing activity before movement, as if they began with a high stimuli-like activity, and then let pre-motor zones take the advantage in terms of signal. It's as if we had a stimulus response, leading to a motor response, but here we must recall that we focus on spontaneous activity: that's why these results are quit surprising, or can let us think about a stimulus-like activation that would reproduce memorized neural process learned during induced activity.

Projection of neural activity on the first two components

We projected neural activity during preparation and inert times on both components. The figure on the left should result on a mean pattern that is exactly the temporal pattern exhibited by the PCA.

An interesting result is the 'regularity' of the temporal patterns: they all seem to follow a same basic pattern with modulation in the amplitude. Further analyses of the neuronal trajectory in the PC space are required.

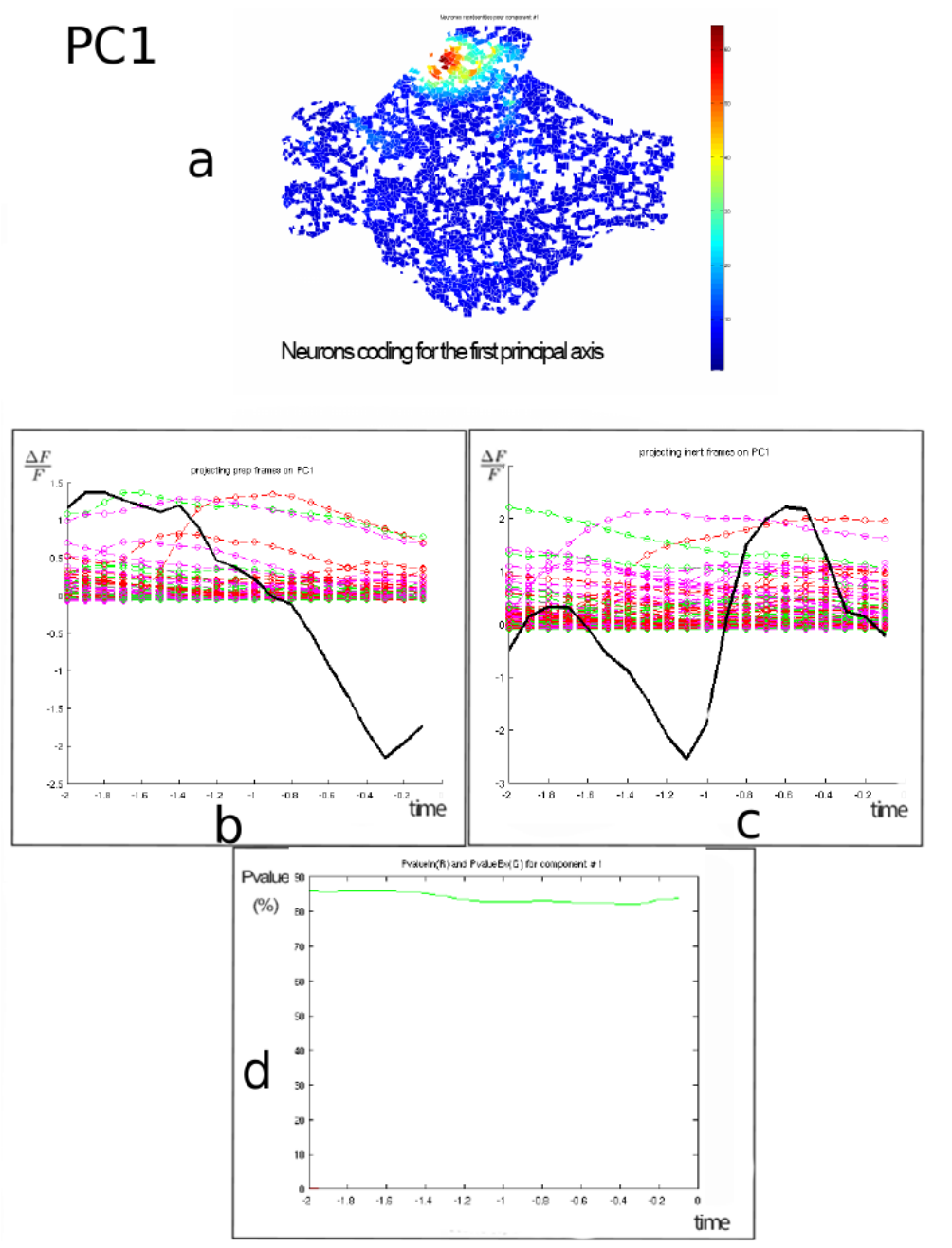

Figure 18: a) map associated with component 1. b) projection of prep frames on component 1. c) projections of inert frames on component 1. d) Mann-Whitney test : we tested if the distribution of the score of PC 1 at a given time before movement was different from its value during inert time.

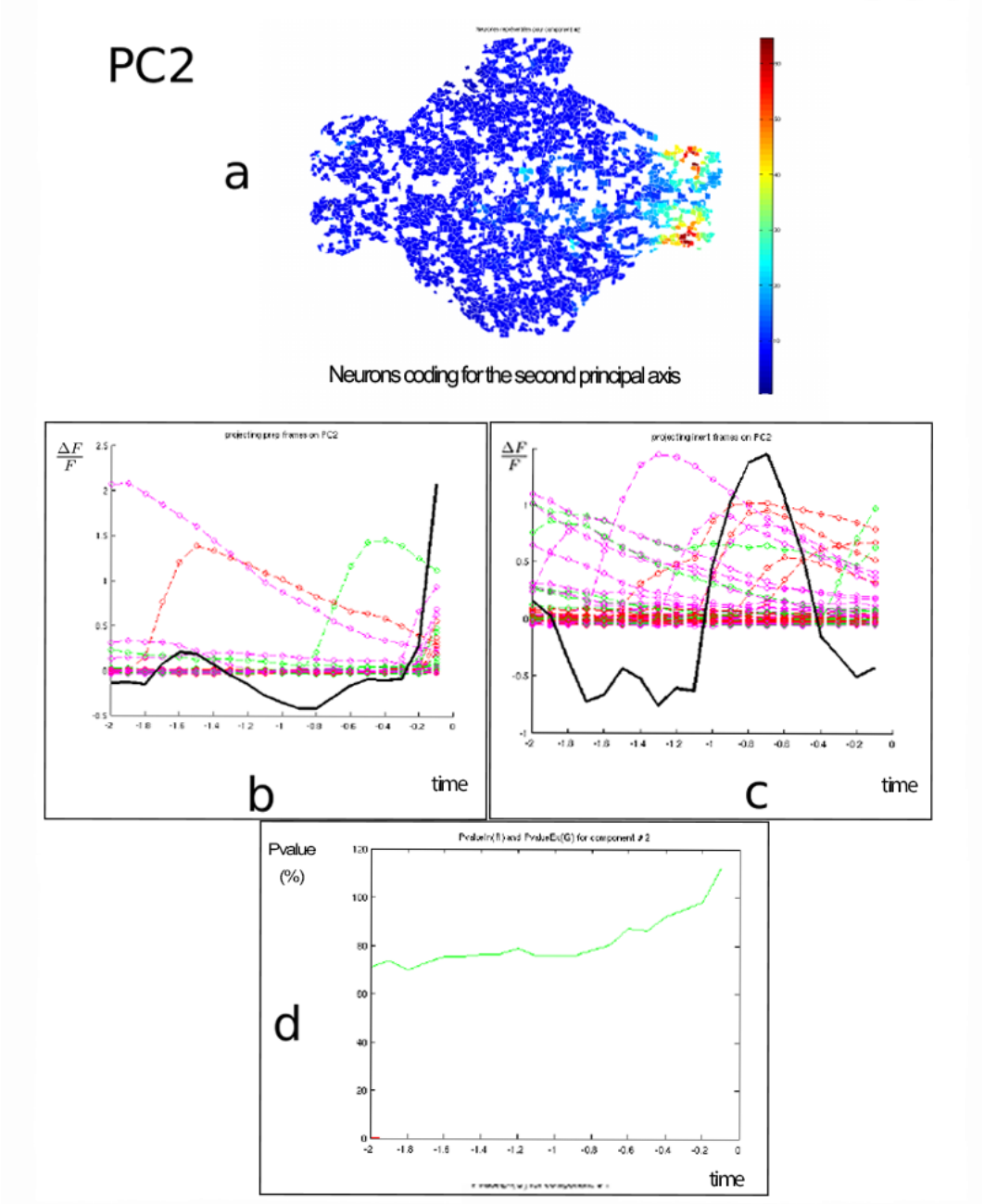

Figure 19: a) map associated with component 2. b) projection of prep frames on component 2. c) projections of inert frames on component 2. d) Mann-Whitney test : we tested if the distribution of the score of PC2 at a given time before movement was different from its value during inert time.

<u>Statistical test on PCA</u>

Thanks to a Mann-Whitney test, we tested if the distribution of the score of PC 1 at a given time before movement was different from its value during inert time.

The results are given in Fig.18.d and Fig.19.d.

We clearly see that neural activity associated with both components is significantly related to preparation of movement, since test results exhibit a same range of differences.

<u>Choice of covariance matrix</u>

A question that we first eluded but that worth been asked is the choice of the correlation matrix. Prospective results do confirm the relevance of the method we have chosen.

Indeed, it is first unclear what frames should be used to compute the covariance matrix for the PCA. That's why we used the eigenvalues obtained after computing the covariance on preparatory frames or inert frames to compare the variance explained by the first PCs in both situations.

We compute :

- Covariance matrix for prep:
  $S_{prep}$
- The associated eigenvectors:
  $u_{i_{prep}}$ for $i \in [1 \cdots N_{comp}]$
- Covariance matrix for inert:
  $S_{inert}$
- The associated eigenvectors:
  $u_{i_{inert}}$ for $i \in [1 \cdots N_{comp}]$

I plotted below the associated confusion coefficients (as defined in [5] by C.J. Machens).

- (Fig.20)

$$CC_{Prep_{prep}} = \frac{u_{i_{prep}}^T . S_{prep} . u_{i_{prep}}}{tr(S_{prep})} \text{ (in blue)}$$

and

$$CC_{Prep_{inert}} = \frac{u_{i_{inert}}^T . S_{prep} . u_{i_{inert}}}{tr(S_{prep})} \text{ (in red)}$$

- (Fig.21)

$$CC_{Inert_{inert}} = \frac{u_{i_{inert}}^T . S_{inert} . u_{i_{inert}}}{tr(S_{inert})} \text{ (in blue)}$$

and

$$CC_{Inert_{prep}} = \frac{u_{i_{prep}}^T . S_{inert} . u_{i_{prep}}}{tr(S_{inert})} \text{ (in red)}$$

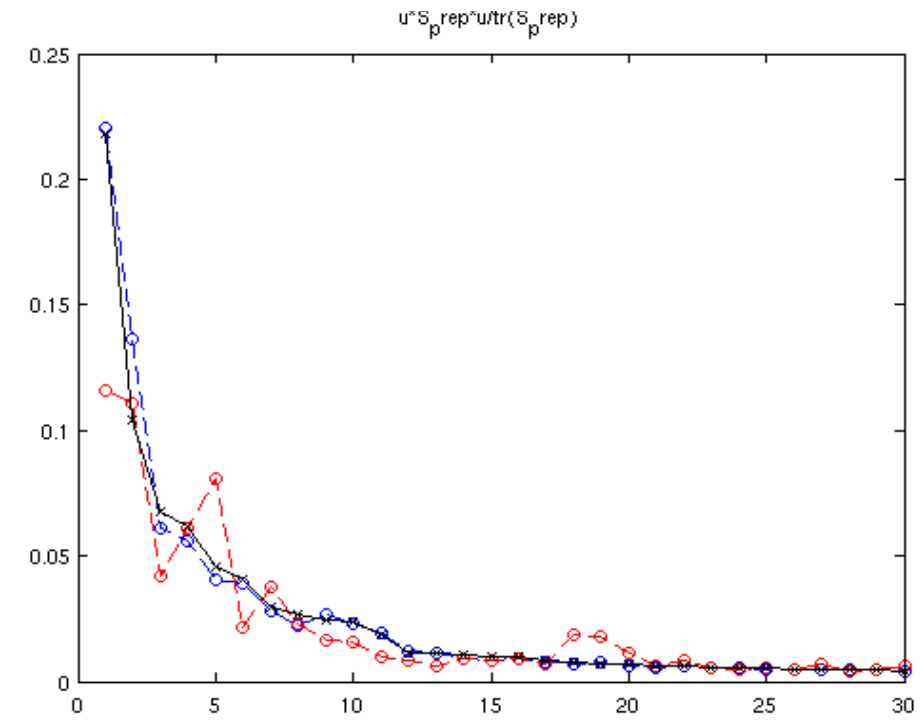

Figure 20: Confusion coefficients for covariance matrix computed on prep frames.

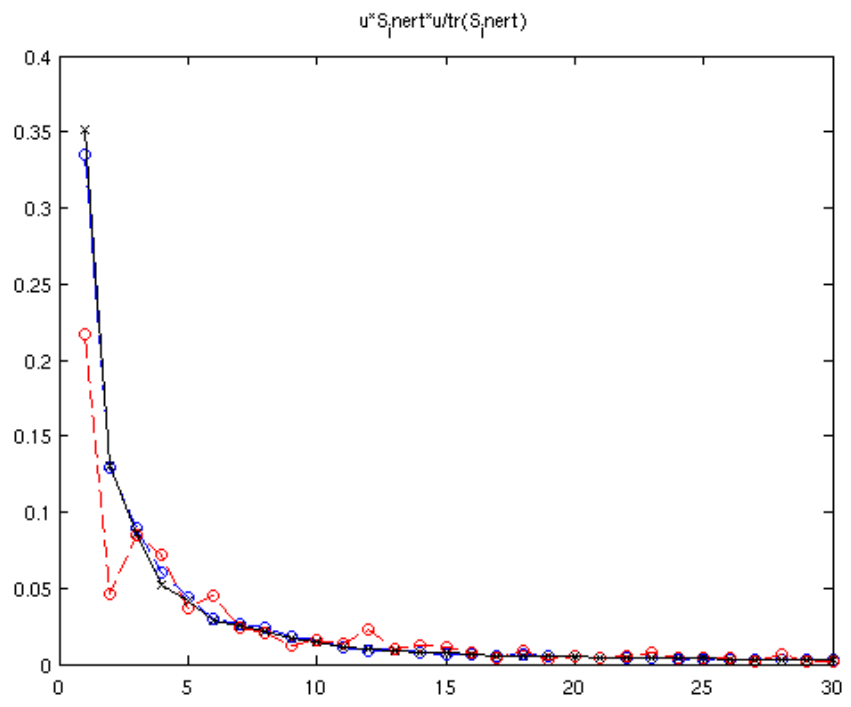

Figure 21: Confusion coefficients for covariance matrix computed on inert frames.

We then plotted the cumulative curves, which correspond to the given of explained variances for each one of these covariance matrices.

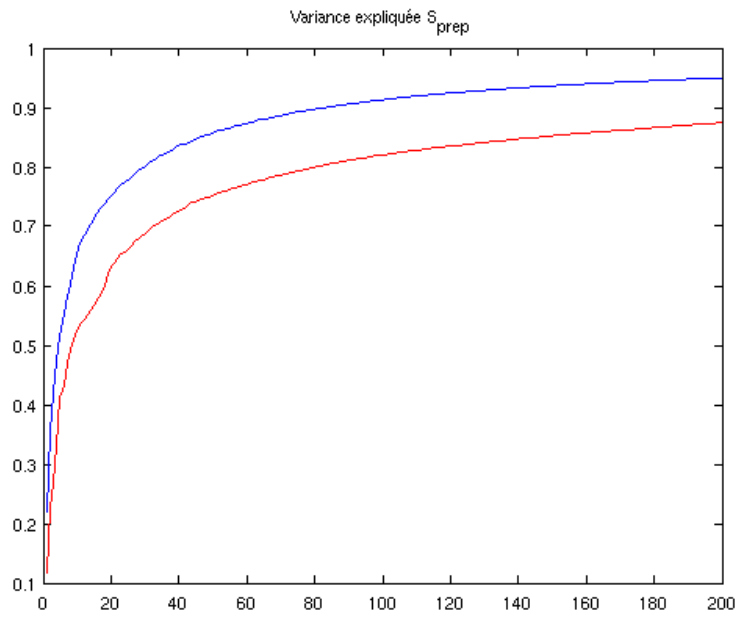

Figure 22: Explained variance for Prep covariance matrix.

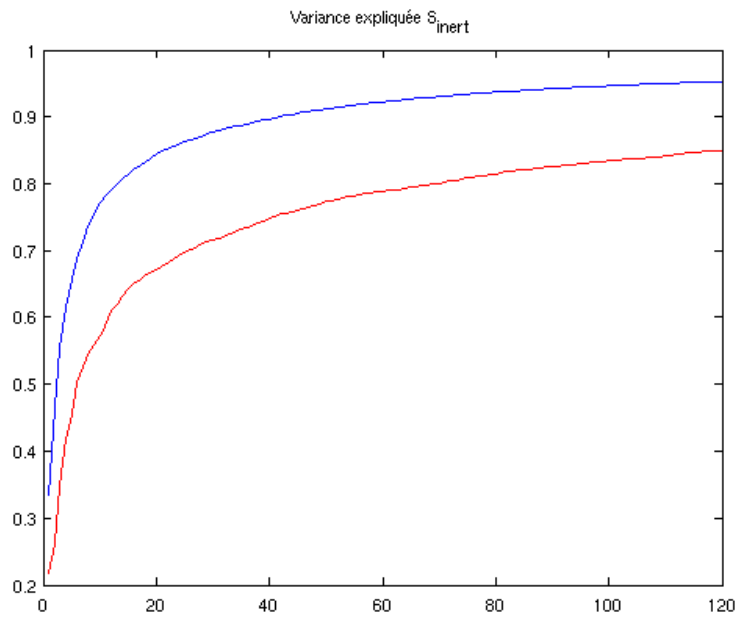

Figure 23: Explained variance for Inert covariance matrix.

The main difference is observable in the first components: they are more discriminant in terms of explained variance for $S_{prep}$.
See for example the difference between red curve and blue curve for the 20th value: the difference is 0.2 for $S_{prep}$, 0.12 for $S_{inert}$, so the case $S_{prep}$ is more discriminative[1].

The cumulative curves (explained variance) are more relevant to figure out how the spaces spanned by these vectors are different.

As a conclusion, we can see that our choice of covariance matrix (we worked with $S_{prep}$ in previous work) was truly relevant given the discrimination between prep times and inert times allowed by this choice.

## III.2.2 Promax

### A relevant clusterization tool

Promax is a method developed by S.Romano et al. (Spontaneous Neuronal Network Dynamics Reveal Circuit's Functional Adaptations for Behavior, 2015) [7]: it exhibits co-activated clusters of neurons combining PCA, rotations and thresholding allowing to define assemblies of neurons and associated co-activated neural patterns.

The framework uses the PCA but adds a rotation (promax rotation) that allows to 'group' or agglutinate coordinates around zero so that it is then easy to select the point subspace by subspace (each subspace is spanned by a limited number of rotated PC) with a threshold that allows to keep points that are in a same proximity area.

[1] *Il y a erreur : soit dans les figures via une permutation Sinert Sprep, soit dans leur lecture -ce qui serait amplement plus dramatique. Les données qui permettraient un recoupement, et donc de trancher, ne sont pas disponibles au moment de la vérification.*

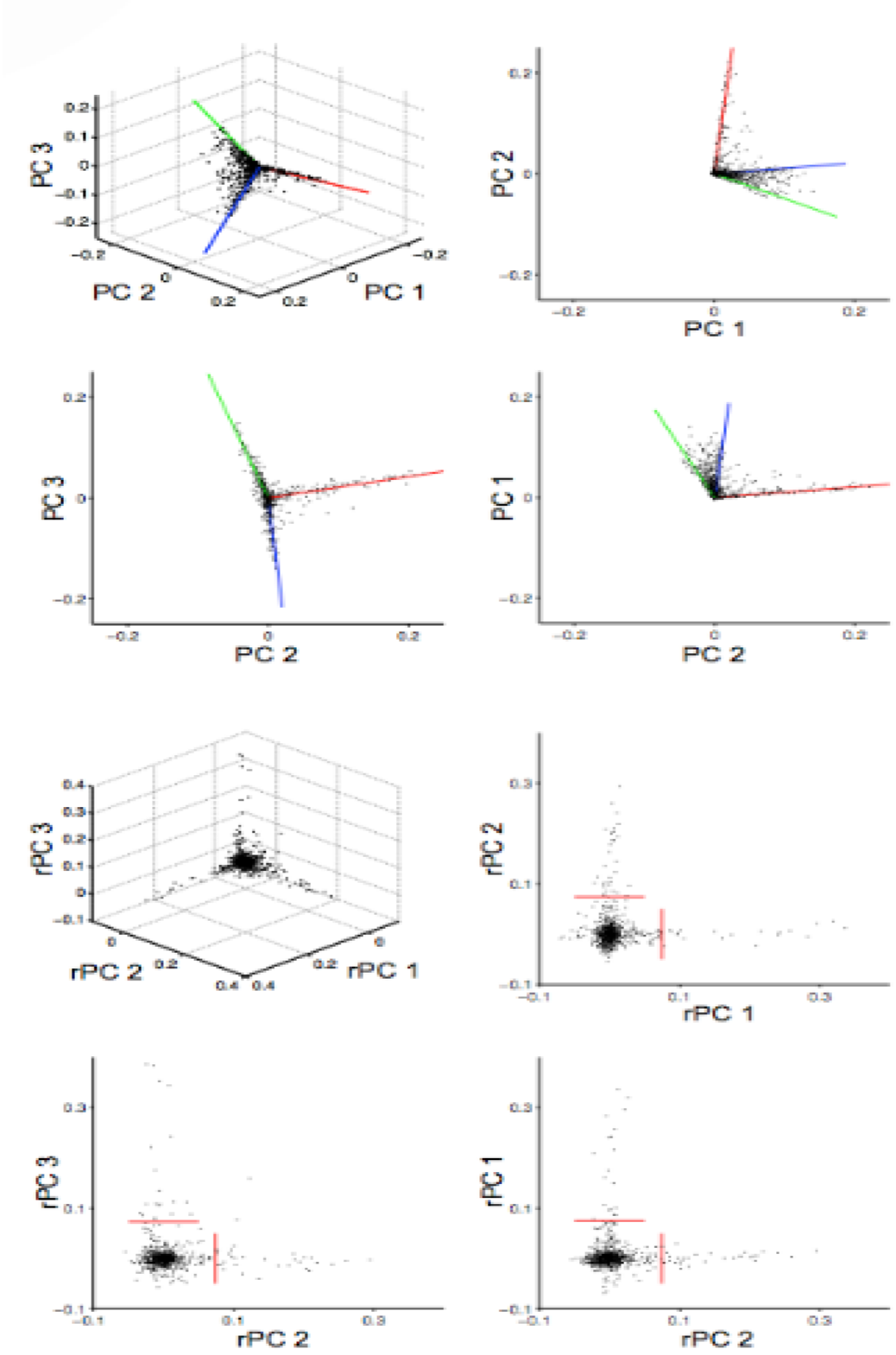

Figure 24: Illustration of promax framework, taken from S.Romano et al. article (Neuron, 2015) [7].

Thanks to this method, we found different groups of neurons that globally co-activate together. Then, to study temporal dynamics associated with these different groups of neurons, we will apply a more time-focus study mask by mask.

Components found by Promax method

I applied the method modulo few qualitative adjustments and a consequent code production. I plotted the results on maps, superposing ROI framed image on real video capture, as usual. I hope that the chosen range of colors allows to represent the diversity of found assemblies.

On our *data*, this method gave the following results:

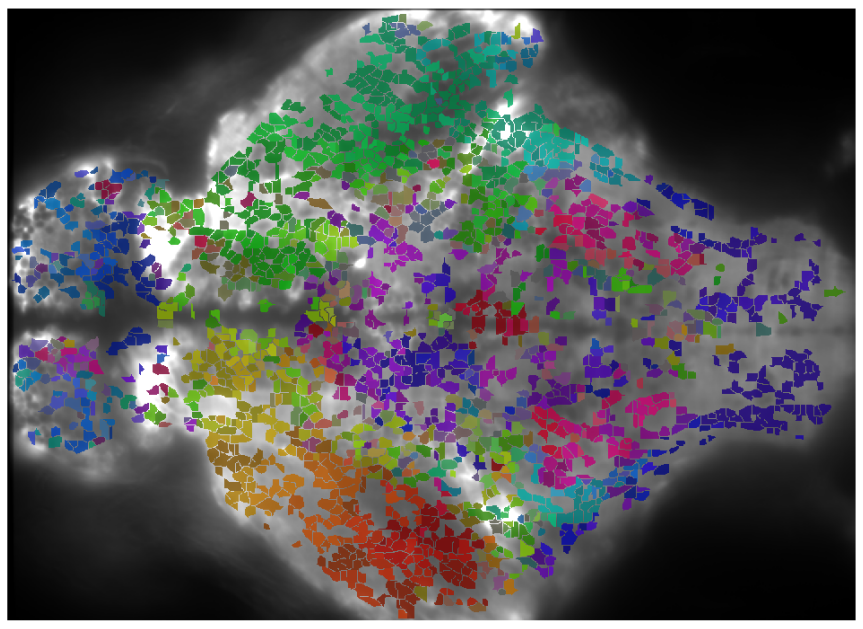

Figure 25: Assemblies found by promax on zebrafish *data.*

Here is an example of one particular assembly, with good symmetrical properties :

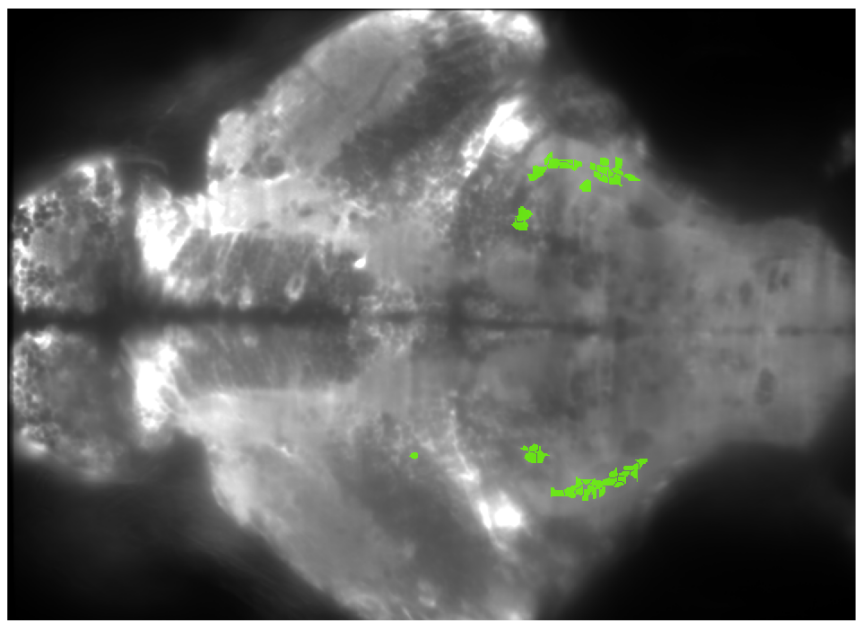

Figure 26: An example of particular assembly.

III.2.3 Searchlight on Promax clusters

Once we obtained the groups of neurons defined by Promax method, we wanted to explore further the associated temporal dynamics and, but, the study of labeled movement's prediction will come later, to see if there was some movement-type specificity associated with each one of these clusters.

Our first try consists in applying Searchlight method on the averaged raster before movement. We want to see how prediction score evolves from group to group, to have a first insight of what are the zones found by the Promax that could be more interesting, and what is the order of magnitude of the score depending on the triggered zone.

Dynamical Searchlight on Promax clusters : why and how?

We slightly change the framework we took before combining the Searchlight and the Promax results, by this time applying searchlight time by time before movement, that means that we look at the score evolution as we come closer to the movement occurrence : the idea is to make the previously explained method run on small time windows (of 300 ms) rolling from 100 up to 1500 ms before movement.

The goal is to see if we find again the ramps exhibited by our first explorations, or if some other patterns are found. The main difference is that here we work with an abstract value, the score, that tells us about the 'power of prediction' of the zone we look at, no more with something homogeneous to $\frac{\Delta F}{F}$ as we intended to do with the PCA or the ICA –formally, however, $\frac{\Delta F}{F}$ is homogeneous to 1 but the 'homoegeneity quality' of the ration is here important.

We slightly change the framework we took before combining the Searchlight and the Promax results, by this time applying searchlight time by time before movement, that means that we look at the score evolution as we come closer to the movement occurrence: the idea is to make the previously explained method run on small time windows (of 300 ms) rolling from 100 up to 1500 ms before movement.
The goal is to see if we find again the ramps exhibited by our first explorations, or if some other patterns are found.

The main difference is that here we work with an abstract value, the score, that tells us about the 'power of prediction' of the zone we look at, no more with something homogeneous to $\frac{\Delta F}{F}$ as we intended to do with the PCA or the ICA.

[illegible]ts

- Searchlight on averaged raster before movement

Here is a table of the scores we get assembly per assembly.

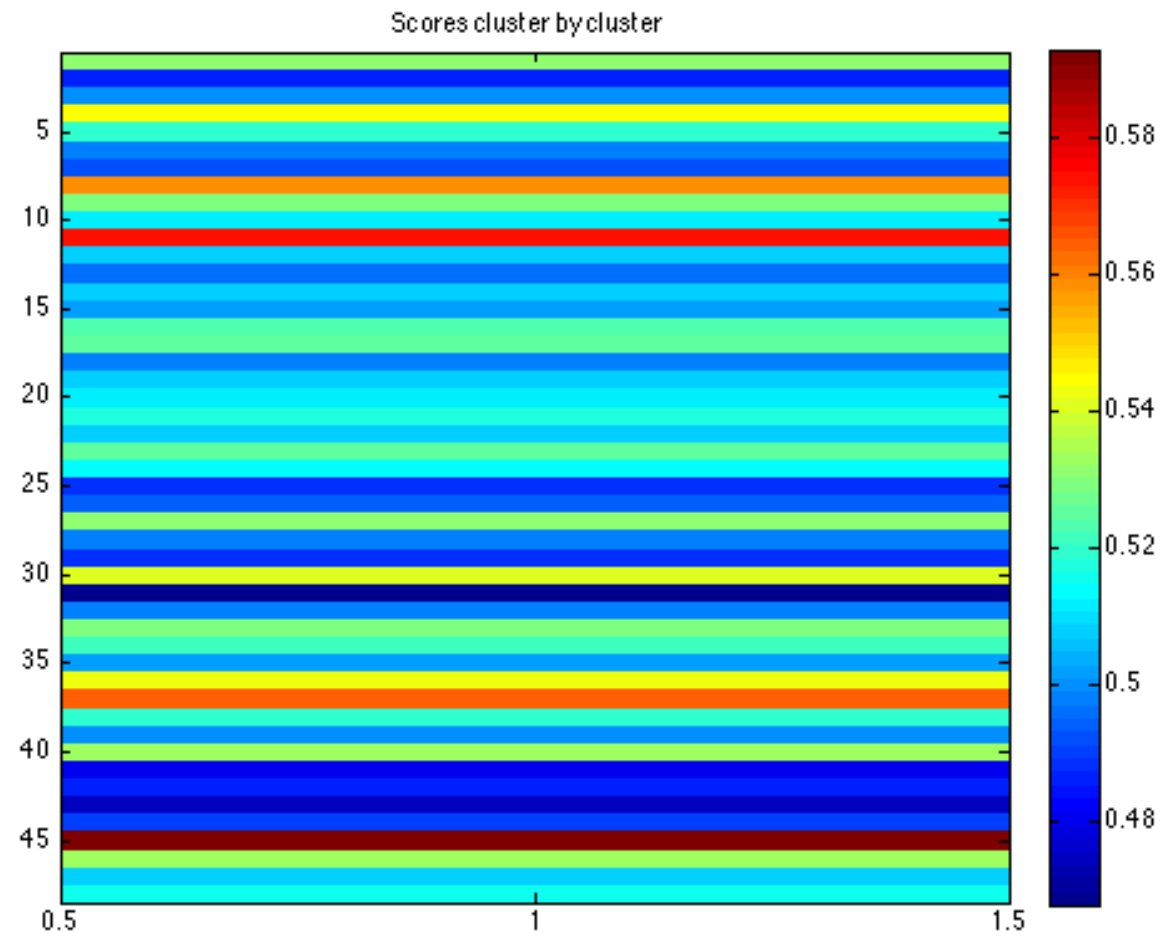

Figure 27: Scores of the searchlight assembly per assembly.

We are not particularly satisfied with these results, since they show great variability in the mean scores and no details about the dynamics of the prediction. In fact, we obviously need here to go further in the temporal aspect of neural activity to see if something relevant can be found by the searchlight.

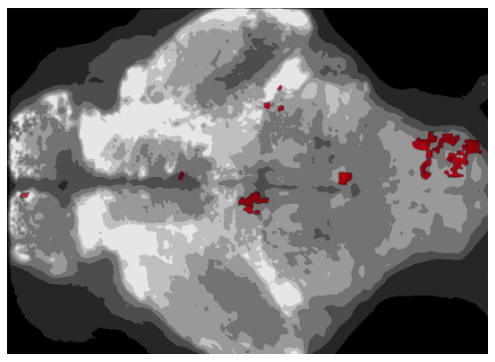

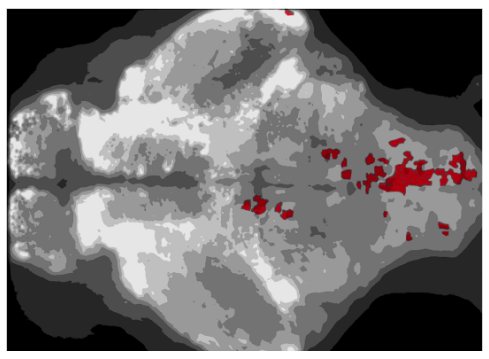

Figure 28: Promax assembly 45
Figure 29: Promax assembly 11

- Dynamical Searchlight

I explained the why and the how of this idea of dynamical searchlight in the previous section. This is a new method, proposed during this internship as something to experiment without any further idea of what it could give exactly. So, I beg the reader to excuse its potential awkwardnesses. Below, the results for dynamical searchlight, from t = 100 ms up to t = 1500 ms before movement(s).

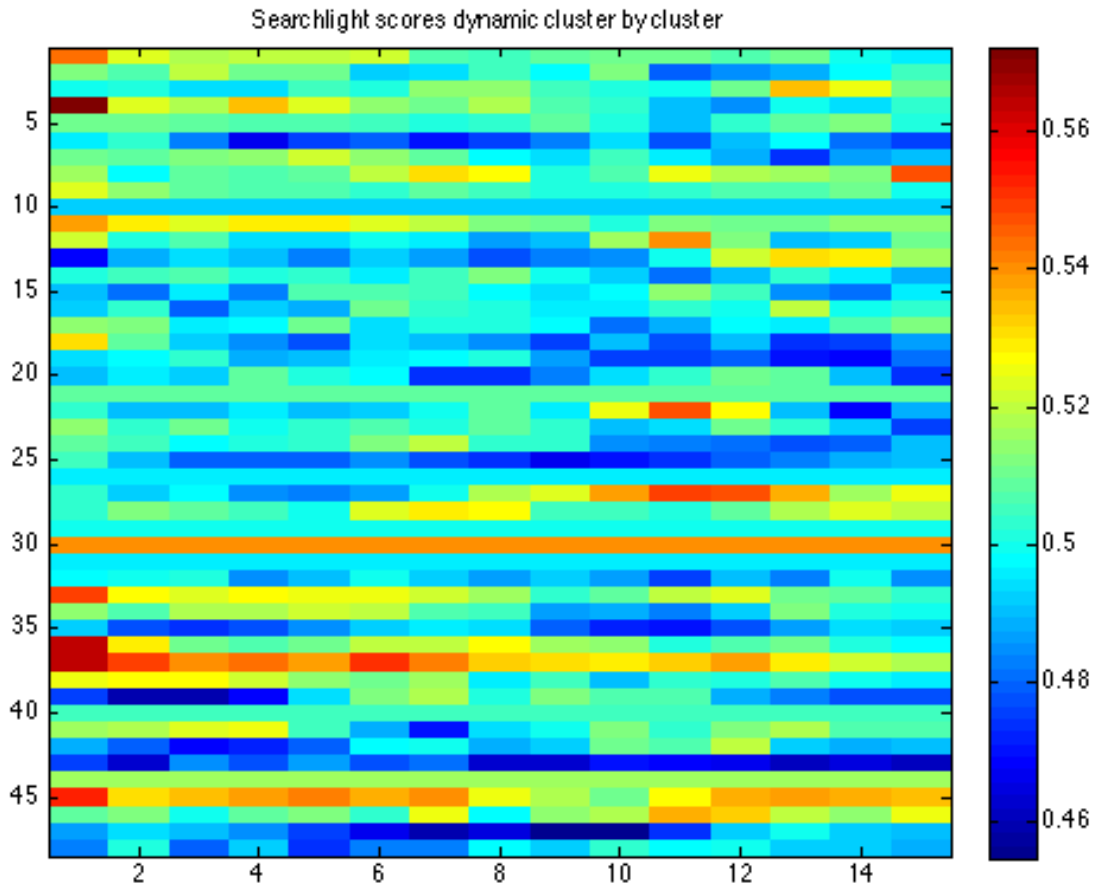

Figure 30: Scores of the dynamical searchlight assembly per assembly -one assembly per line.

What's more, assemblies that exhibit best scores correspond to zones mostly marked in pre-motor areas (Fig.28 and Fig.29).

For six of these assemblies, we notice a last moment increase of prediction ability (for example see assemblies 1, 4, 11, 33, 36, 37 or 45). This increase is true for other assemblies, but less remarkable.

For some of them, prediction peak is achieved far before the movement, 1200ms before movement approximately.
See for example assembly 22 or 27. The case of assembly 37 is interesting since it activates sooner that the other assemblies that exhibit pre-movement firing.
And when we map this assembly, one can clearly see that it contains neurons that do not belong to immediate pre-motor area (Fig.31).

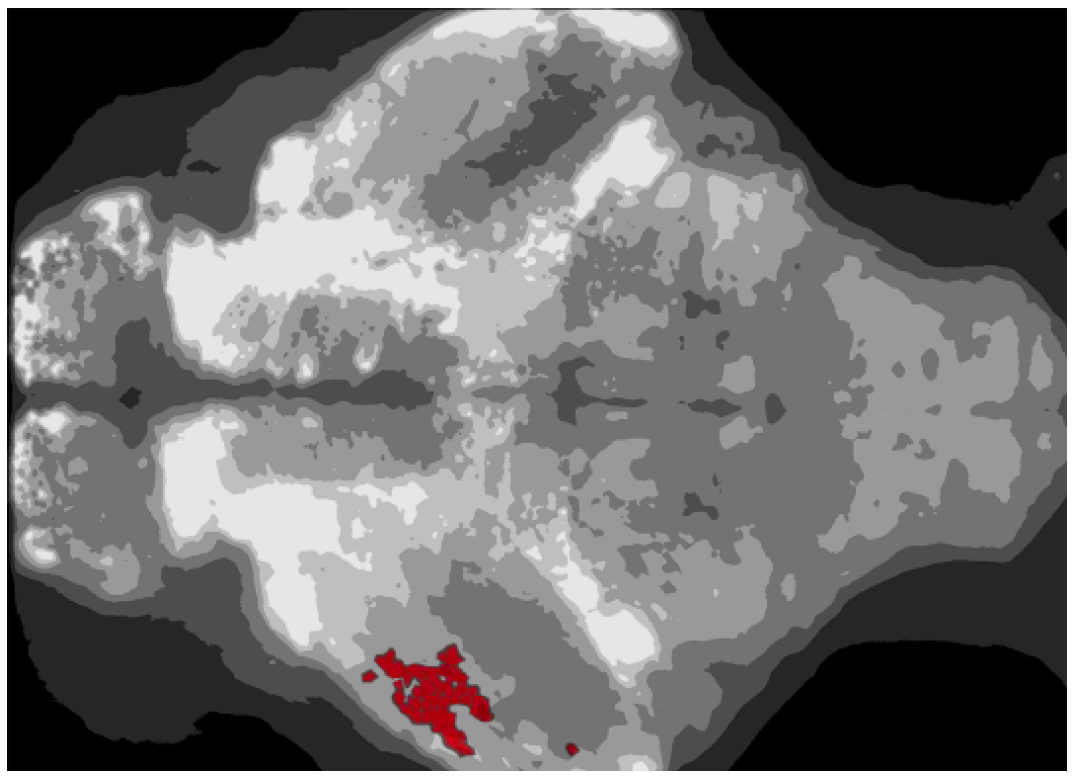

Figure 31: Mask 37 shows a slowly growing score results up to a maximum of prediction at time t = 100ms before movement. That could tend to show that neurons close to sensory areas are involved in spontaneous motor activity generation.

Concerning the up and down peak of prediction (that's how I chose to call these pretty good scores achieved far before the movement and then decreasing down to a normal base line, such as for assembly 22 or 27), here are the maps of neural zones associated with these dynamics:

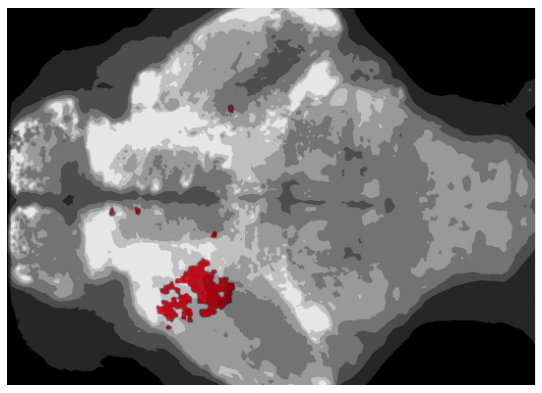

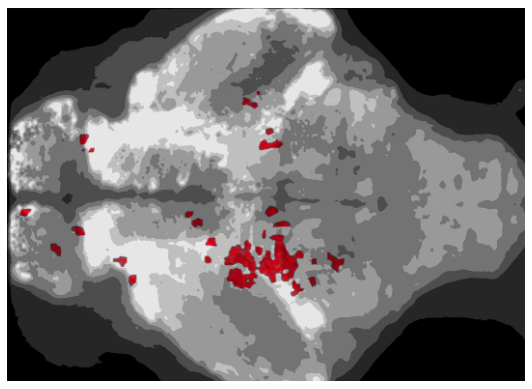

Figure 32: Promax assembly 22
Figure 33: Promax assembly 27
Figure 34: {Fig.32,Fig.33} Masks 22 and 27 show an up and down prediction ability through time. I chose to show associated maps since it refers to zones that the searchlight had shed light on in previous results.

<u>III.2.4 Direct Linear Discriminant Analysis</u>

Linear Discriminant Analysis separates two classes maximizing the inter-classes scatter and minimizing the within-classes scatter, working on the ratio $\frac{\sigma_{between}}{\sigma_{within}}$ where the σ are the empirical variances (in fact we work with scatter matrices, finding projection matrix A such that $\frac{A.S_b.A^T}{A.S_w.A^T}$ is maximal where $S_w$ is within-class scatter matrix and $S_b$ is between-classes scatter matrix. A should diagonalize both $S_w$ and $S_b$, with: $A.S_w.A^T = I$ and $A.S_b.A^T = \Lambda$, where Λ is diagonal. As zeros in $S_b$ do not carry any significant information, whereas zeros on $S_w$ correspond to information allowing to discriminate classes, the idea is to introduce a step that keep only non-zeros values of the diagonalization $V.S_b.V^T= \Lambda$, calling Y the (first) *m* non-null columns of $V.Y.S_b.Y^T=D_b$ where $D_b$ contains the first *m* elements of Λ (that means that $D_b$ is the $m$x$m$ principal submatrix of Λ).
Then, writing $Z = Y.D_b^{-1/2}$, it appears that Z unitizes $S_b$ (*i.e.* $Z.S_b.Z^T = I_m$).

Now that a first step is achieved, and reminding that we want to find a matrix that diagonalizes both $S_b$ and $S_w$, the idea is to introduce U such that U diagonalizes $Z^T.S_w.Z$[2] :
*i.e* s.t. $U.(Z^T.S_w.Z).U^T=D_w$, where $D_w$ is diagonal.

Now, defining the matrix A by $A = U^T.Z^T$ we have :

[2] Note that we also reduce the dimensionality of $S_w$, but not by truncation.

$$A.S_w.A^T = D_w \quad \text{and} \quad A.S_b.A^T = I \; .$$

The job is done then, at a normalization step close (left-multiplying the *data* by $D_w^{-1/2}$.A in order to sphere them).

A figure presented in the original direct-LDA article [8] proposed by H.Yu and J.Yang sums up well all these steps :

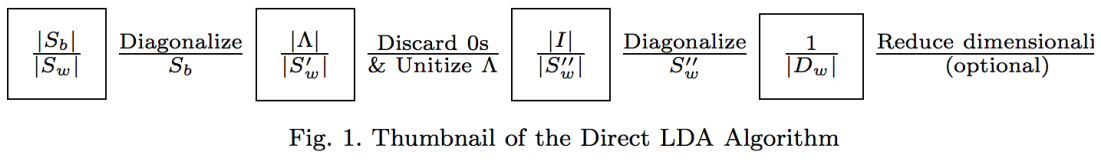

Figure 35: DirectLDA framework

Then, looking at the classification score performed by direct-LDA implementation on our *data*, we have a tool that we can use similarly to Searchlight, with the same initial framework (we perform direct-LDA dynamically, that means that we make the computation run time step per time step from 3000ms before movement up to 100ms before movement with a step of 100ms).

<u>Maps</u>

Here are the results at particular times, where zones involved reminded regions that we already had guessed as involved in movement preparation. Here the task is limited to disciminating pre-movement *vs* inert frames:

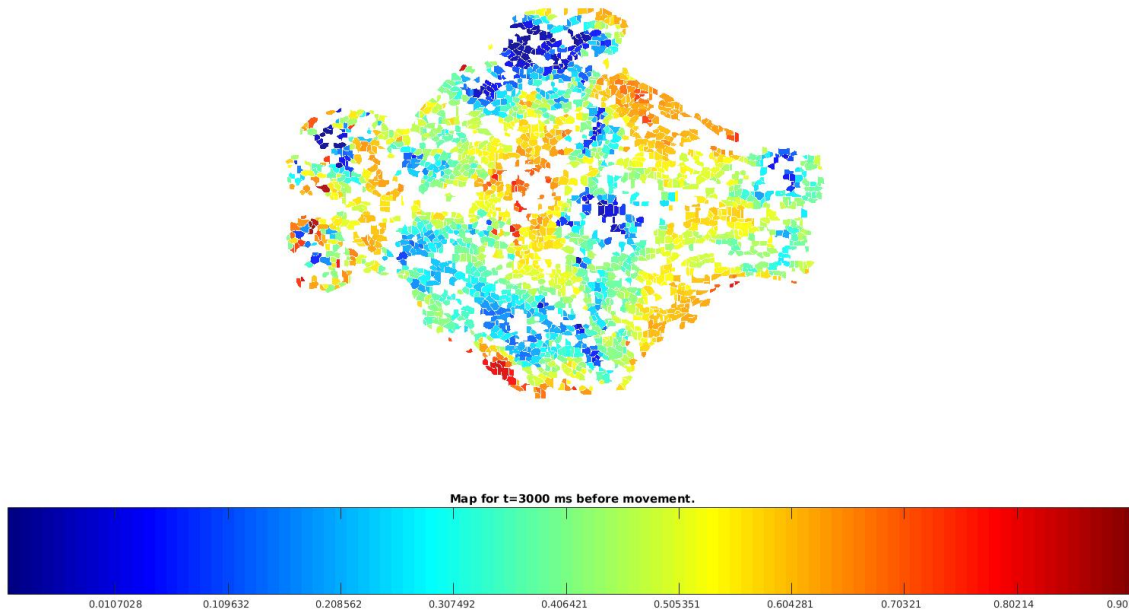

Figure 36: Scores map for Direct LDA 3 seconds before movement occurrence

We can see that far before the movement, the only zone significantly discriminative is a region we had seen with the first searchlight map, and that is close to optic tectum. That could confirm that areas close to sensory regions are involved in spontaneous motor activity generation. It is a region close to the one presented in Fig.32 (Promax assembly 22) -we had reported an up and down score results for this zone, with dynamical searchlight method (svm score)- but sharper in this case.

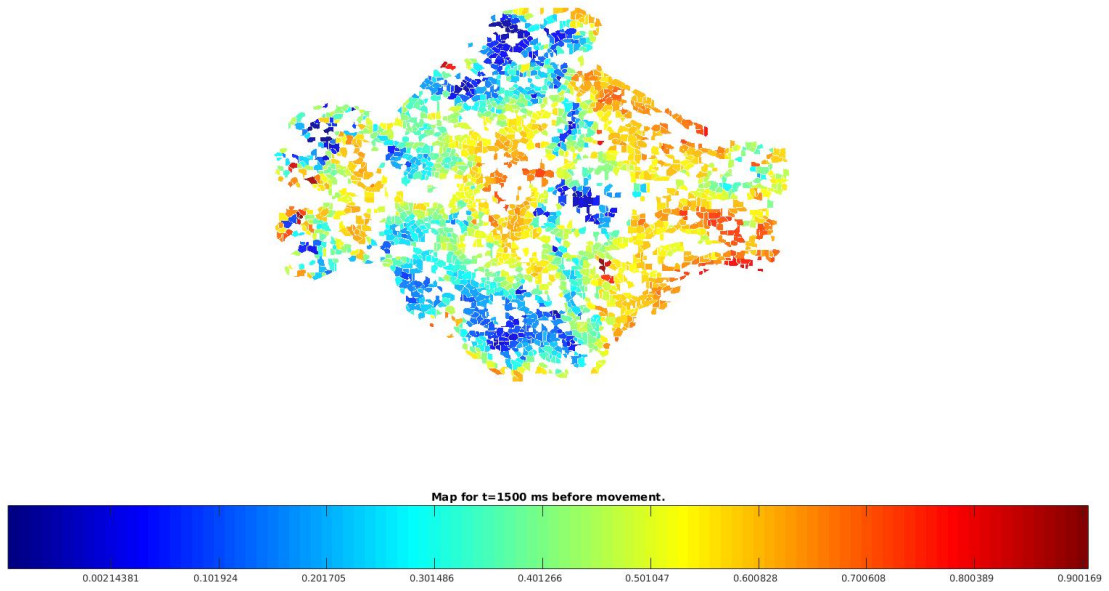

Figure 37: Scores map for Direct LDA 1.5 seconds before movement occurrence

Now, 1.5sec before movement, we can see in Fig.37 that this region is no longer fired and that pre-motor areas are associated with good score. But then this zone becomes good for prediction again, almost as good as the whole pre-motor region (Fig.38, Fig.39), as another region, close to optic tectum, exhibits surprisingly good scores too just before movement.

I have no explanation to propose for this last moment prediction pick concerning this zone close to the optic tectum, which had already appeared in the first bank of analysis we had performed. Coupling these observation with precise biological knowledge of these zones could obvisouly lead to a proposition. Let's recall the possibility of pattern-like coactivations, which have no function but testify for an usual concomittance of activation(s) in the case of a stimulus provoked movements.

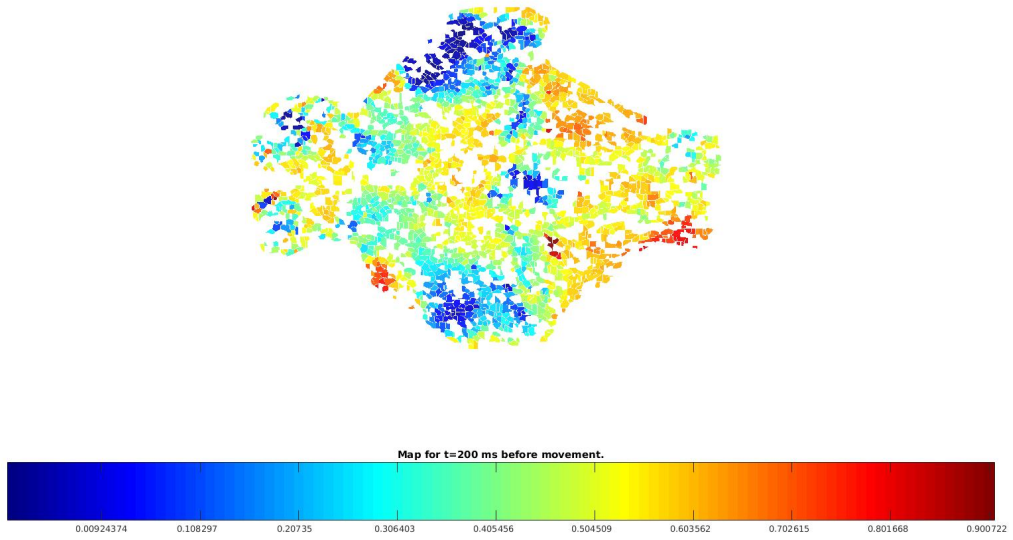

Figure 38: Scores map for Direct LDA 0.2 seconds before movement occurrence

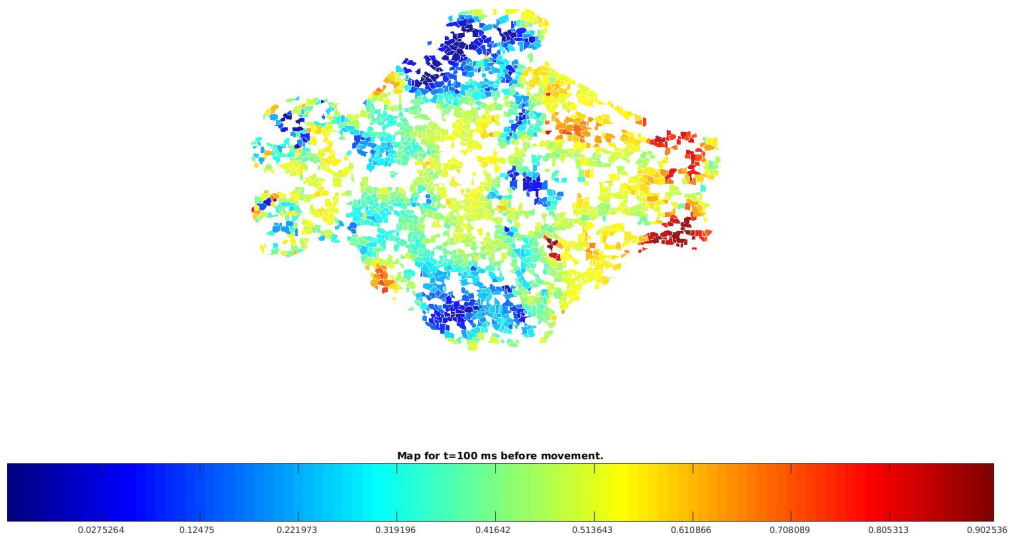

Figure 39: Scores map for Direct LDA 0.1 seconds before movement occurrence

## III.2.5 Independent Components Analysis

A suggested approach was to try Independent Component Analysis in order to extract groups of independent co-activated neurons (that lead to spatial patterns). To perform an ICA, we can choose the number of components (or we can let it be equal to $N_{neurons}$) and we need to choose a function used to estimate the negentropy. We found that changing the number of components had little effect whereas change in the estimator of negentropy (in fact change in the contrast function) had tremendous influence on the component found. We couldn't find meaningful pattern of activation.

### Results

We first chose the contrast function canonically proposed:

• $G_1(u) = \frac{1}{a_1}.\log(\cosh(a_1 u))$ with $1 \le a_1 \le 2$. This choice is known as a good general purpose contrast function (fig34).

Then, we tried:

• $G_2(u) = -\frac{1}{a_2} e^{a_2.\frac{u^2}{2}}$, which is usually used for "highly super gaussian *data*", or when robustness is very important (fig35).

and

• $G_3(u) = \frac{1}{4} u^4$ , used only for estimating subGaussian IC (fig36).

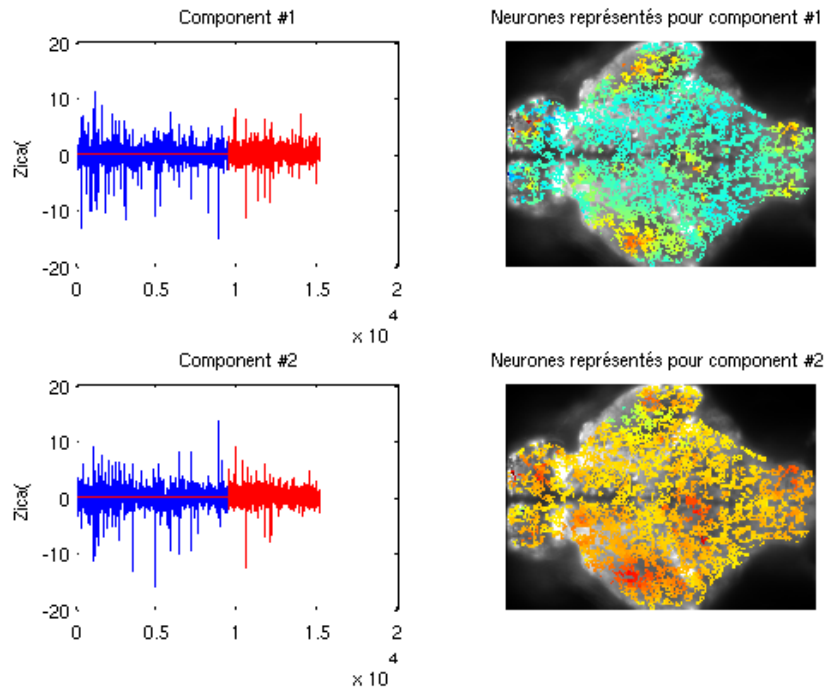

Figure 40: First components and their associated maps for contrast function $G_1(u) = a.\log(\cosh(a.u))$

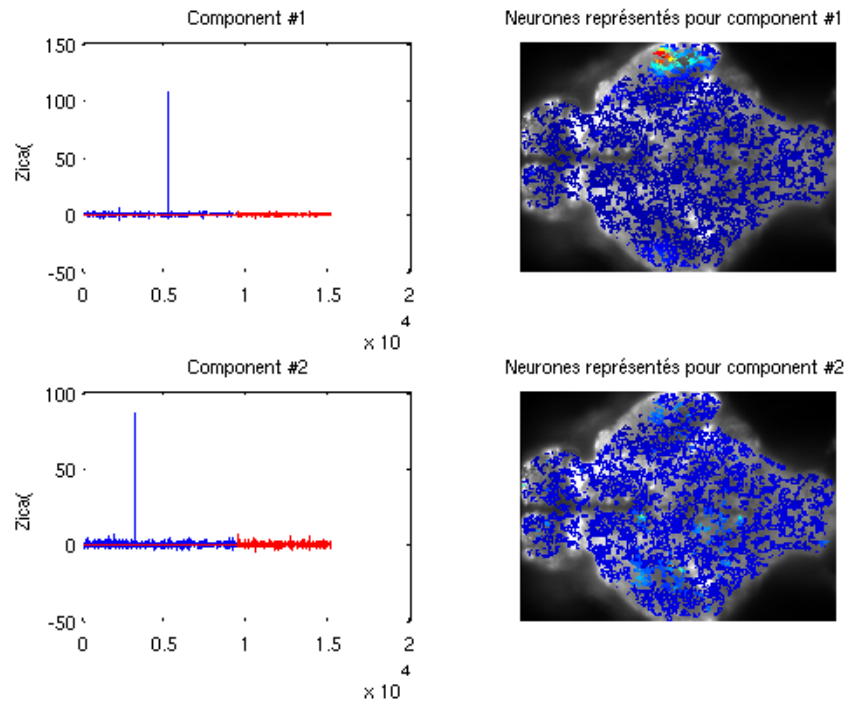

Figure 41: $G_2(u) = 1/a.\exp( a.u^2/2)$

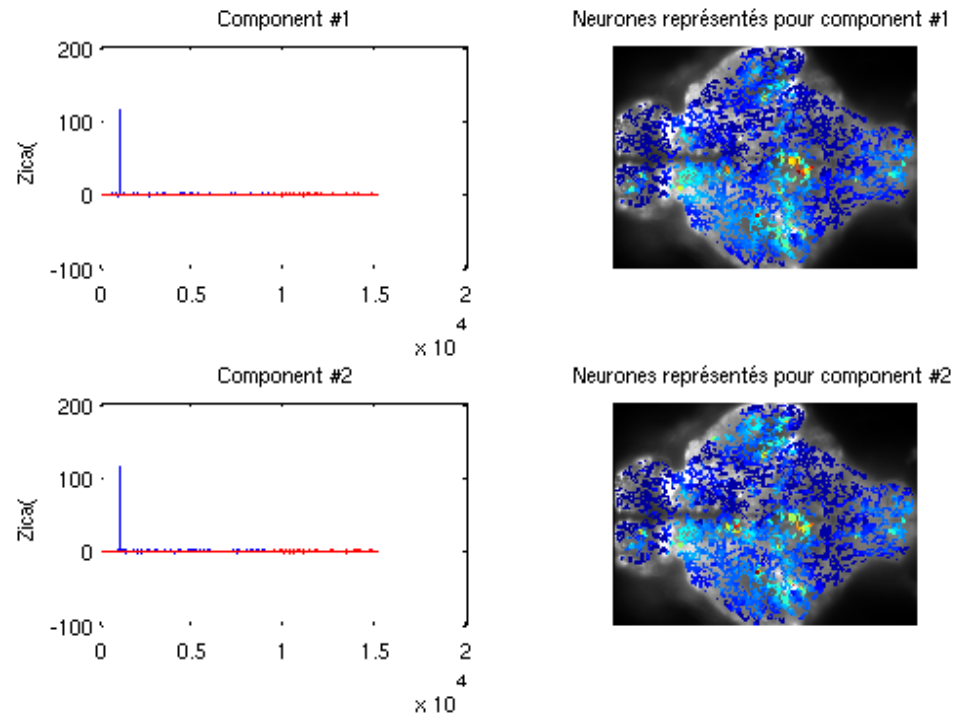

Figure 42: First components and their associated maps for contrast function $G_3(u) = 1/4.u^4$

### Discussion

For $G_1$ we get component activations and maps that lack sparseness. We expected more convincing results with $G_2$, but the results shown in fig35 and fig36 make us notice that, this time, we have spatial map interesting but the temporal activations associated with these maps are too sparse.

They correspond to one activation or co-activated neurons at one time.

We got the same type of 'movie-decomposition' with $G_3$ : the ICA run the decomposition too far, we get one map per time step, as if the only way to decompose the neural video *data* into independent components was to go back to each image that compose the movie.

As a conclusion, we couldn't get meaningful pattern from the ICA, this method might have failed because independent patterns cannot be found in a recurrently connected network, in fact the hypothesis of independence is too strong, as a complex temporal process occurs with potentially inter-playing zones : exhibiting independent activation patterns would rely on this unrealistic hypothesis of independence.

## III.3 Dynamics of prediction for labeled *data*

III.3.1 Framework

The assumption underlying his work is that we can somehow write Mvt = f ($\overrightarrow{r_{t-\tau}}$), where $\tau$ goes from 100ms up to 1500ms. As a first approach, and after having looked at the histograms associated with the different categorizations of movement that we can make, I chose to work with two categories that correspond to big movements *vs* small movements. Indeed, if you look to the plot of tail movement associated with the different clusters found by Promax method, one can notice that, choosing the good threshold, there is a pretty clear distinction between small amplitudes and big amplitudes that allows to split them in two distinct categories.

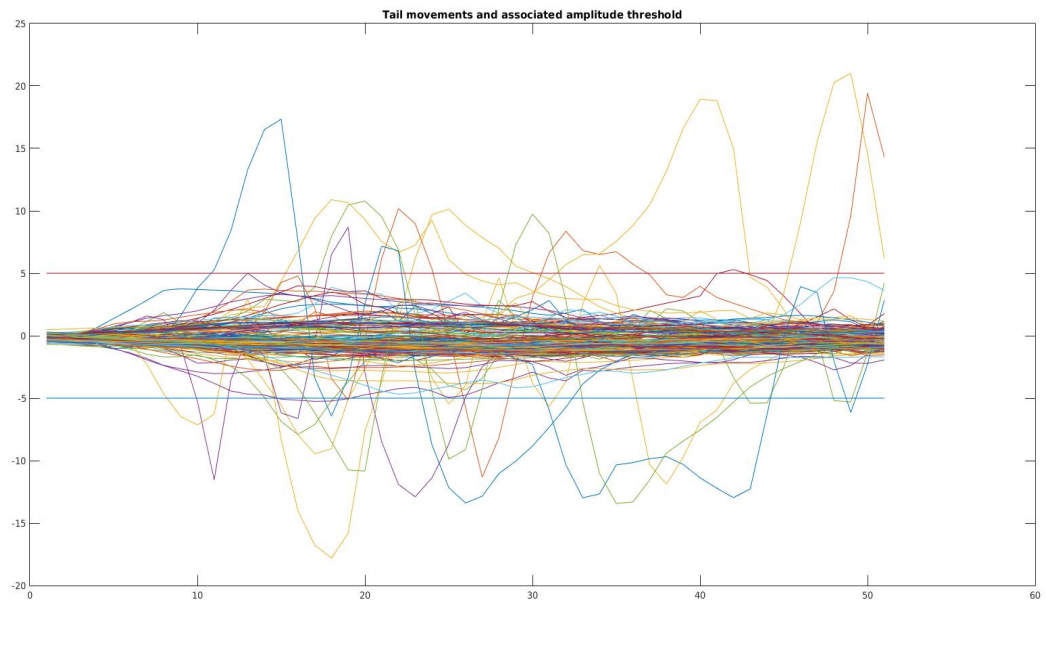

Figure 43: Tail movements and threshold for classification.

What's more, as usual some frames are labeled as 'Inert' frames -they correspond to the case where there is no movement in a wide enough window containing this frame. So we finally have three kinds of frames we will work with:

A={Prepframes corresponding to big movement}

B={Prepframes corresponding to Small movement}

C={Inertframes}

Each frame is represented, for the svm, as a point in a 30-dimension space (corresponding to the associated activity rate on each neuron, *i.e.* each axis. So the svm has to classify this point as a frame belonging to one or the other category. This two categories choice can be:

- A *vs* B (score $s_1$)
- A *vs* Inert (score $s_2$)
- B *vs* Inert (score $s_3$)

III.3.2 Dynamical searchlight on Promax clusters

We can distinguish five possible cases of separability of the classes by the SVM used in searchlight, that, as we use a linear kernel, are directly linked with the geometry of the *data*. The combination of the svm scores associated with each one of these tests can helps us to understand how *data* is organized and which one of these hypotheses about the geometrical organization of *data* is verified.

- h1 A B Inert non-distinguishable
- h2 A B *vs* Inert
- h3 A *vs* B Inert
- h4 B *vs* A Inert
- h5 A *vs* B *vs* Inert

Then, we can make a preferred assumption about how specific to one type of movement (A? B? None of them?) or not a cluster found by the Promax method can be.

Let's call $s_1$ A *vs* B svm score. Let's call this score $s_2$ for A *vs* Inert classification, and $s_3$ for Inert *vs* B classification.

- h1: $s_1$ $s_2$,$s_3$.
- h2: $s_2$,$s_3$ $s_1$.
- h3: $s_1$,$s_2$ $s_3$.
- h4: $s_1$,$s_3$ $s_2$.
- h5: $s_1$,$s_2$,$s_3$.

| | A | B |
|---|---|---|
| B | $s_1$ | |
| Inert | $s_2$ | $s_3$ |

*Table 1: SVM Scores as an insight of data geometry.*

III.3.3 Results

Pieces of results, in themselves

In Fig 44, 45, 46, we plotted the dynamical prediction score for each assembly exhibited by Promax method (each line corresponds to one assembly), each column to time from -100ms up to -300ms before movement.

In all the cases we can see that prediction becomes better as we come closer to the movement, but this is more evident for the discrimination between Big Amp and Small Amp.

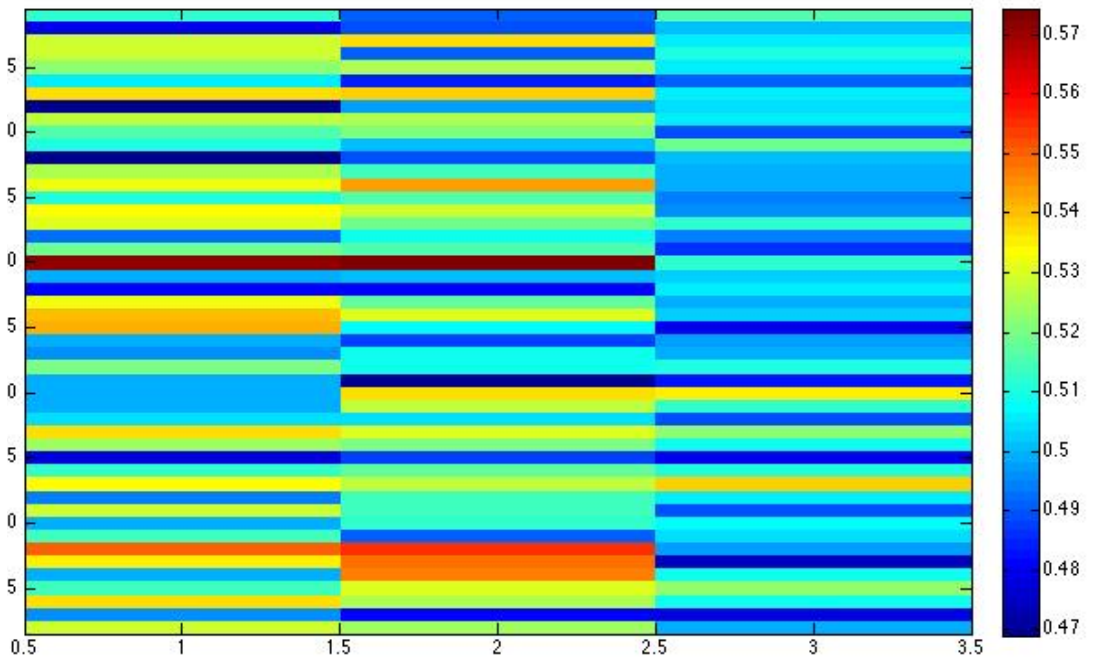

Figure 44: Scores $s_1$ for all assemblies (Big Amp *vs* Small Amp tested time by time (x axis), for each one of the promax assemblies (y axis: one line refers to an assembly)).

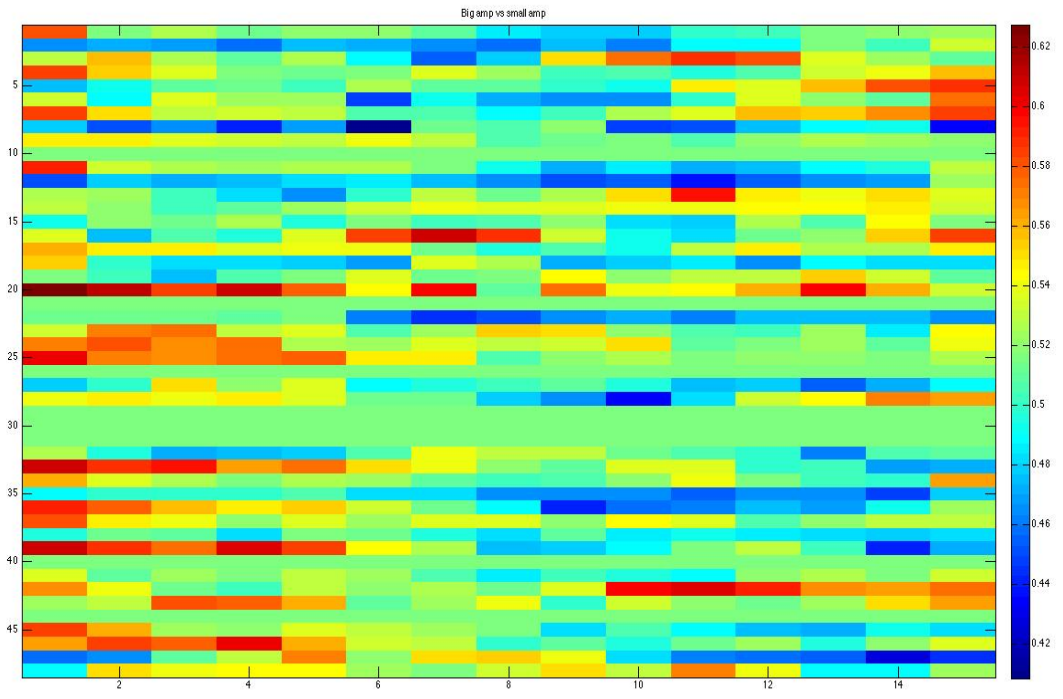

Figure 45: Scores $s_2$ for each one of the assemblies (Big Amp *vs* Inert tested time by time (x axis), for each one of the promax assemblies (y axis: one line refers to one assembly)).

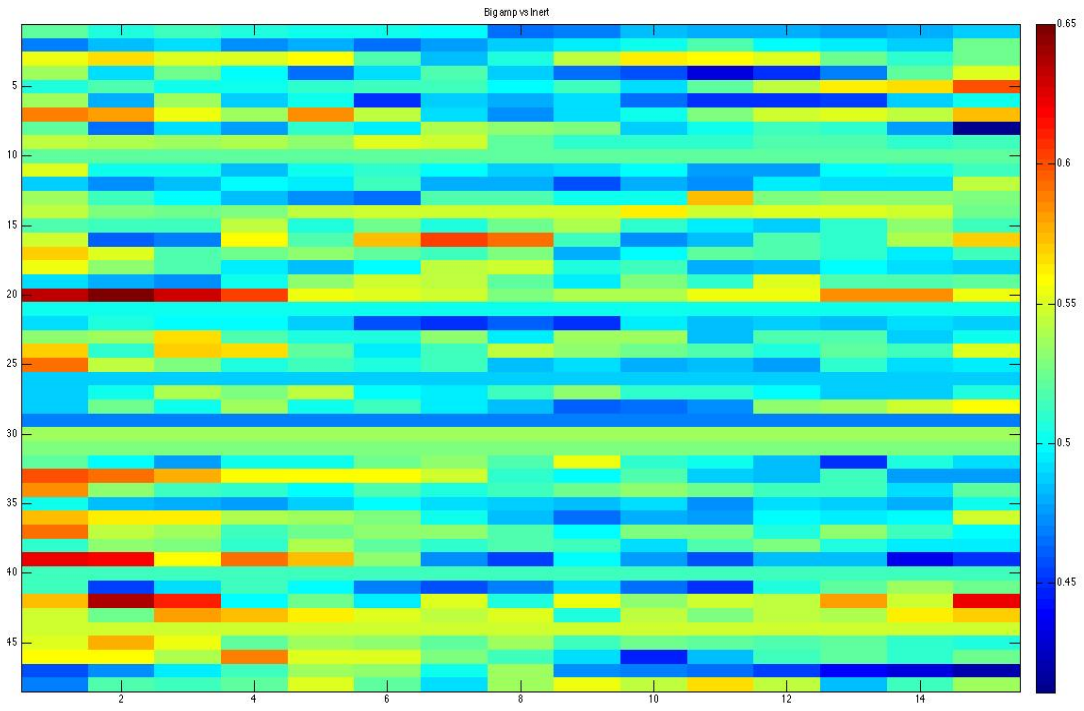

Figure 46: Scores $s_3$ for all assemblies (Small Amp *vs* Inert tested time by time (x axis), for each one of the promax assemblies (y axis: one line refers to one assembly)).

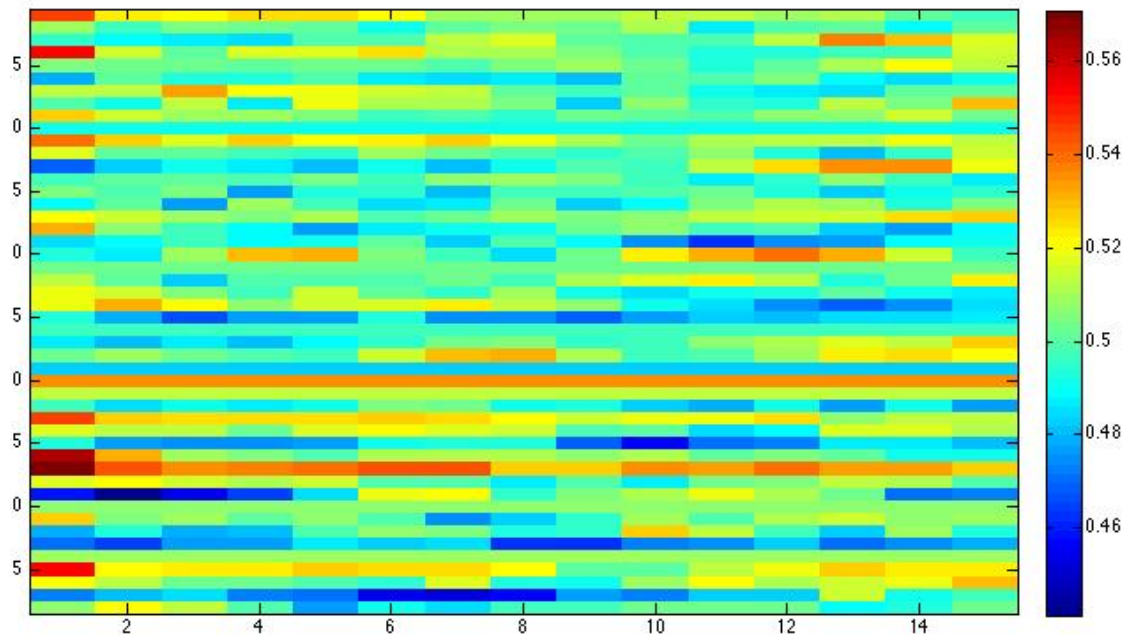

Figure 47: Scoretable. Each line refers to an assembly, $s_1$ $s_2$ and $s_3$ are presented from left to right.

This score table gives an idea of how movements can be easily discriminated in function of associated preparation neural activity: we see there that big movements are well discriminated from small movements and from inert times.But discrimination between small movements and inert times does not lead to good results. According to the hypotheses framework we proposed, $h_3$ ($s_1$, $s_2$ $s_3$) is validated.

We can then make two assumptions.

-1- Neural activity associated with a movement is 'proportionally' linked to his amplitude, that means that for small movement neural activity is not easily distinguishable from noise. The idea that big movement are easily distinguished from inert times that small one is nevertheless not as surprising as another result could have been, since it confirms what first intuition could have tell us.

-2- Promax assemblies were computed thanks to a method that use a PCA computed on prep frames correlation matrix. Since we know that,

we can assume that these zones are specifically activated before a movement and that the given of these zone distribution do not contain the information associated with inert frames. That would mean that a bias would exist in the choice of the neurons we look at, not in the geometry of the neural *data* itself.

<u>Comparing with classical searchlight results</u>

First we can notice that we find again last moment increases of prediction ability for assemblies 1, 4, 11, 33, 36, or 45 for all set of discrimination hypothesis $s_i$.

We can look at some assemblies that share a same ramp score pattern for each one of the $s_i$ ($i \in [1,3]$) score representation presented previously in the tables.

- For $s_1$ (A *vs* B, *i.e* discrimination between Big Amplitude movement preparation and Small Amplitude movement preparation), assemblies showing remarkable score increase are: 1, 4, 11, 33, 36, 39, 45, 46 (let's call $\text{Assemblies}_{s_1}$ this ensemble).

- For $s_2$ (A *vs* Inert, *i.e* discrimination between Big Amplitude movement preparation and Inert frames): 1, 4, 11, 33, 36, 45 (let's call $\text{Assemblies}_{s_2}$ this ensemble).

- For $s_3$ (B *vs* Inert, *i.e* discrimination between Small Amplitude movement preparation and Inert frames): 20, 33, 34, 36, 37, 39 (let's call $\text{Assemblies}_{s_3}$ this ensemble).

We notice that :

$\text{Assemblies}_{s_2} \subset \text{Assemblies}$

$\text{Assemblies}_{s_1} \cap \text{Assemblies}_{s_2} \cap \text{Assemblies}_{s_3} \neq \varnothing$

These results are coherent with non-labeled dynamical searchlight best assembly -the one that presents a ramp like score across time. Globally, it correspond to the union of the set of assemblies represented here for $s_1$, $s_2$ and $s_3$.

So results are coherent.

# IV. Conclusion

The very notion of causality is not trivial, nor even usable in this context if it had only depended on a wise and non contingently-rushed opinion -which is precisely not an opinion, I guess.
However, I hope that the reader has allowed him(/her)self to feel free to understand it as a 'Humeian' would have, when we spoke about correlative schemes, or as a more *a priori* 'Kantian' notion when a need for more "high of sight" did reveal itself as necessary –see for example the book of Michel Malherbes, "*Qu'est-ce que la causalité ? Hume et Kant*", coll. *Pré-textes* pub. house *J.Vrin*.

This internship was the opportunity to explore many investigation methods for a both high dimensional (number of descriptive variables for each elements) and large (number of elements) *data* set of neural *data*. All of these methods were not presented in this report because six months is a short delay to fully apply each one of the encountered or imagined prospection tools, and the labeled *data* came pretty late in the schedule (the whole set-up had been to be done again, with a longer time of observation: initially the fish was observed two hours during, the experience protocol has been changed for an observation time fixed to six hours, in order to have more movements, and then to have a sufficient representation of each label -please see A. Jouary thesis work for details).

From April to September, statistical descriptive approaches like t-test or Mann-Whitney test were tested and mapped, after having customized all the pre-processing code material allowing to organize the *data*; then came the searchlight as a first raw approach for mapping an intuition of the zones that could be interesting to study for behavior prediction. Quickly, the idea of the PCA came out as a central tool for finding both temporal patterns and associated maps, since it allows at the same time to decrease problem dimensionality and to exhibit a new basis of expression of the neural *data* that we can easily interpretate spacially speaking thanks to mapping and in terms of signal thanks to the associated components. Such an advantage came from the linearity of PCA method, that exhibit some vectors homogeneous to the initial variables (we had some neurons, and some time signals for each neurons, PCA exhibited linear combination of neurons, and linear combination of time signals). Then, the ICA is a natural try since it relies on the same main approach excepting that a stronger assumption is made (not only non correlation, but independance): but as we explained, this assumption of independance is non realistic since we work with *data* recorded in a network with a lot of crossed influences. The PCA-Promax approach proposed by S.Romano et al allowed us to implement another approach of the problem, looking for groups of neurons coactivated, *i.e.* synchronized in time in their activations. The dynamics exhibited were numerous and sometimes really similar, so this method was not relevant for exhibing time dynamics but rather to define neurons clusters that present a coherent group-like activity. We decided to keep these clusters to look further to their intrinsec dynamics applying other set of tools in a dynamical way. That's why we decided not only to question to what extend we could adapt the searchlight to implement a dynamical approach of the problem –making the discrimination between pre-movement activity and inert activity run time by time, and showing how prediction score evolves up to the movement-, but also to make this dynamical analysis run cluster by cluster. The direct-LDA came as a last technical more directly readable than the searchlight, and gave some results globally coherent with what we had got before (pre-motor area exhibit ramp-like score, the same region close to tectum showed an interesting arize too). The reason why we can keep direct LDA as a more relevant method than the Searchlight is the way that it calls an algorithm more adapted to high dimensional *data* than the classical svm implemented in the searchlight (as we said, the problem of over-fitting was a reef to constantly keep in mind and to avoid). What's more, implementing searchlight took a lot of time, since we had to choose the good kernel, to look for the good number of folds for the cross validation, *etc.*: all these exploration steps are not necessary in LDA method.

Many ideas came during these months that still remain to explore. I think for example to bi-clustering adapted to a tensorial representation of the *data*, which would be a new way to modelize the problem solving both neural and temporal approach in a same computation [10].

Furthermore, we thought a lot about the assumption we can make about the functional role we could imagine for the neurons: we could distinguish state neurons from decision neurons. For representation and coding convenance, not necessarily real neurons but 'abstract' neurons constructed by linear combination of existing biological neural entities. State neurons would be always in activity in a slot-like rate, and the state (high or low) at time t would decide of the lateralization of any potential movement that would occur at this time. The occurrence time of a movement would be decided by the so called 'decision neuron'. Such a general modelization of the problem still remains to explore, and has been questionned by Adrien Jouary (IBENS, zebrafish neuro-ethology) in his PhD thesis. As a conclusion, let's remember that we showed that neural activity was significantly (and/notably qualitatively) different before movement not only in pre-motor areas, and that the specific activation patterns were linked with these activation times. At the end, we showed that these patterns of activation were linked with the score prediction any learning algorithm can make, so these increases of the activity can be indeed been supputated as taking part in the causality chain of spontaneous movements. Remains the problem of the variability of the results: we can see e.g. the divergence of scores that one can read in promax clusters.

## V. Acknowledgement - References

I warmly thank Adrien Jouary (IBENS, PhD student) for having been a motivating, demanding and comprehensive tutor during this internship, and Jonathan Boulanger Weill (IBENS, PhD Student) for his well-thought support.

Sebastian Romano, formerly post-doc in the lab, helped me a lot to see how encouraging PCA results could be, and explaining me how we could relevantly use the Promax method he had developed last year in this same lab.

Alexandre Gramfort for his early monitoring : comments and reading advices helped a lot to shape the searchlight's construction, ICA suggestion enlightened latter understanding of newer methods.

Vincent Lostanlen, PhD student in the Informatics Department of the ENS, for having sugested a tensor representation of the *data*.

I also would like to thank Alessia Candeo for her benevolent support during the end of this internship.

Finally, I obviously thank a lot German Sumbre for his advises and for having encouraged me to enroll for a phD, and Sophie Deneve for having begun to suggest me some readings that could be useful for the next.

I propose below a short bibliography : I decided to present only articles that really provided me methods or suggestion of methods which were implemented during my internship.